%% file: rhicrev_main.tex
\documentstyle{ar209}

\begin{document}

\input epsf.tex    
\input epsf.def    
\input psfig.sty

\jname{Annu. Rev. Nucl. and Part. Phys.}
\jyear{2006}
\jvol{1}
\ARinfo{1056-8700/97/0610-00}

\title{Results from the Relativistic Heavy Ion Collider}

\markboth{BERNDT M\"ULLER and JAMES L. NAGLE}{RHIC RESULTS}

\author{Berndt M\"uller
\affiliation{Department of Physics, Duke University, Durham, NC 27708-0305}
James L. Nagle
\affiliation{Department of Physics, University of Colorado, Boulder, CO 80309-0390}
}


\begin{abstract}
We describe the current status of the heavy ion research program
at the Relativistic Heavy Ion Collider (RHIC). The new suite of 
experiments and the collider energies have opened up new probes 
of the medium created in the collisions.  Our review focuses
on the experimental discoveries to date at RHIC and their interpretation
in the light of our present theoretical understanding of the dynamics
of relativistic heavy ion collisions and of the structure of strongly
interacting matter at high energy density. 
\end{abstract}

\maketitle


\section{The Science of RHIC}

\subsection{Strongly interacting matter at high energy density \label{sect1a}}
\input{section1a.tex}

\subsection{``The Quest for the QGP'' revisited \label{sect1b}}
\input{section1b.tex}

\subsection{Context in physics and astrophysics \label{sect1c}}
\input{section1c.tex}


\section{Overview of experimental results}

\subsubsection{Update on RHIC detectors \label{sect1d}}
\input{section1d.tex}

\subsection{Global characteristics, yields, and spectra \label{sect2a}}
\input{section2a.tex}

\subsection{Flow observables \label{sect2b}}
\input{section2b.tex}

\subsection{Hard observables \label{sect2c}}
\input{section2c.tex}


\section{Interpretation of Results}

\subsection{The "ideal fluid" \label{sect3a}}
\input{section3a.tex}

\subsection{Energy loss jet tomography \label{sect3b}}
\input{section3b.tex}

\subsection{Thermodynamic properties \label{sect3c}}
\input{section3c.tex}

\subsection{Partonic collectivity \label{sect3d}}
\input{section3d.tex}


\section{Outlook}

\subsection{Review of lower energy results \label{sect4a}}
\input{section4a.tex}

\subsection{Future outlook for RHIC \label{sect4c}}
\input{section4c.tex}

\subsection{Outlook for LHC program \label{sect4b}}
\input{section4b.tex}

\subsection{Open theoretical problems \label{sect4d}}
\input{section4d.tex}

\section{Summary \label{sect5}}
\input{section5.tex}

\section{Acknowledgements}
We wish to thank Bill Zajc, Tim Hallman, Wit Busza, and Jean-Paul Blaizot for a careful reading of 
the manuscript and many useful suggestions.  We also thank Frithjof Karsch for providing the data
points in Figure 1.  We acknowledge support from the United States Department of Energy grants
DE-FG02-05ER41367 (M\"uller) and DE-FG02-00ER41152 (Nagle).


\input{biblio.tex}
\end{document}

%% file: section1a.tex

The Relativistic Heavy Ion Collider (RHIC) at Brookhaven National 
Laboratory was constructed to investigate the properties of nuclear 
matter at ultra-high energy densities \cite{Harris:1996zx}. 
The motivation for this research arose in the 1970s, first, from 
speculations about possible novel states of nuclear matter at 
supranuclear densities \cite{Bodmer:1971we,Lee:1974ma}, and soon
afterwards, from the insight that the structure of matter should 
become simple at asymptotically high temperatures \cite{Collins:1974ky}. 
The celebrated principle of {\em asymptotic freedom} states that 
the effective coupling constant $\alpha_s$ of quantum chromodynamics 
(QCD), falls with increasing momentum transfer $q^2$ or, 
equivalently, with decreasing distance between particles. In a 
thermal medium, the characteristic momentum transfer between (nearly) 
massless particles is of order the temperature $T$, and 
thus the effective coupling between quarks and gluons must become 
weak when $T$ grows large. The complicated structure of nuclear 
matter at low temperatures, where it is composed of a multitude of 
hadronic particles, baryons and mesons, whose interactions are 
difficult to calculate with precision, is thus expected to give way 
at high temperatures to the relative simplicity of a plasma composed 
of weakly interacting quarks and gluons, the {\em quark-gluon plasma}
\cite{Shuryak:1977ut,Shuryak:1978ij}.\footnote{Note that 
this argument does not apply to cold (i.e. low temperature) nuclear 
matter at high baryon density, because the Pauli principle forbids
most scattering processes among quarks, except those near the Fermi 
surface. As a result, the structure of asymptotically dense, cold 
quark matter is more complex. See \cite{Alford:2001dt} for a review.}

Numerical solutions of QCD, discussed below, tell us that the transition 
from a hadronic gas to a quark-gluon plasma should occur at a temperature 
of approximately 170 MeV. This value coincides with the ``limiting'' 
temperature of matter composed of hadrons first postulated by Hagedorn 
\cite{Hagedorn:1965st}. The Hagedorn temperature, defined as the 
exponential slope of the mass spectrum of hadronic resonances, was later 
understood to be associated with the transition to a new phase of matter
\cite{Hagedorn:1980kb}, where hadrons are dissolved into their constituents, 
quarks and gluons. The precise value of the transition temperature $T_c$, 
and how high the temperature must rise before the plasma can be 
considered as weakly coupled, can only be determined by accurate, 
nonperturbative simulations of the equations of QCD. 

Over the past decade, such solutions have become available due to vastly improved numerical 
simulations of thermal QCD on a Euclidean space-time lattice. In this 
approach, one evaluates the path integral of the QCD partition function 
\begin{equation}
Z_{\rm QCD} = \int DA \prod_f D\psi_f D\bar\psi_f 
  \exp\left( - \int_0^{1/T} d\tau \int d^3x 
  {\cal L}_{\rm QCD}[A,\psi_f,\bar\psi_f] \right)
\label{eq1a1}
\end{equation}
by a sum over representative field configurations $A(x,\tau)$, 
$\psi_f(x,\tau)$, and $\bar\psi_f(x,\tau)$, where $A$ denotes the gluon
field, $\psi_f, \bar\psi_f$ the quark fields of various flavors $f$ 
and their adjoints, and $\tau$ is the imaginary time. Thermodynamic 
variables are obtained as derivatives of $\ln Z$; thermal expectation 
values of other observables can be obtained by inserting the relevant 
expression for the operator into the path integral. 

The functional integral over the quark fields requires the evaluation 
of the determinant of a large space-time matrix, which is a numerically
demanding task.\footnote{In the so-called {\em quenched approximation} this 
determinant is replaced by a constant. This procedure amounts to the 
neglect of all thermal quark-antiquark excitations, which is generally 
not a good approximation at finite temperature. Results obtained with 
this approximation must be be considered with caution.} The numerical 
expense grows rapidly with decreasing mass of the quarks. Significant 
progress has been made on the evaluation of the integral (\ref{eq1a1}) 
in recent years, mainly due to three innovations: Improvements in computer 
hardware permit much faster calculations (up to about $10^{10}$
floating-point operations per second); improved representations of 
the QCD action on a lattice have enhanced convergence to the continuum 
action; new lattice representations of the quark fields preserve the 
symmetries (chiral symmetry) of QCD in the massless quark limit. 
Calculations using the last improvement are presently under way.

\begin{figure}
\centerline{\psfig{figure=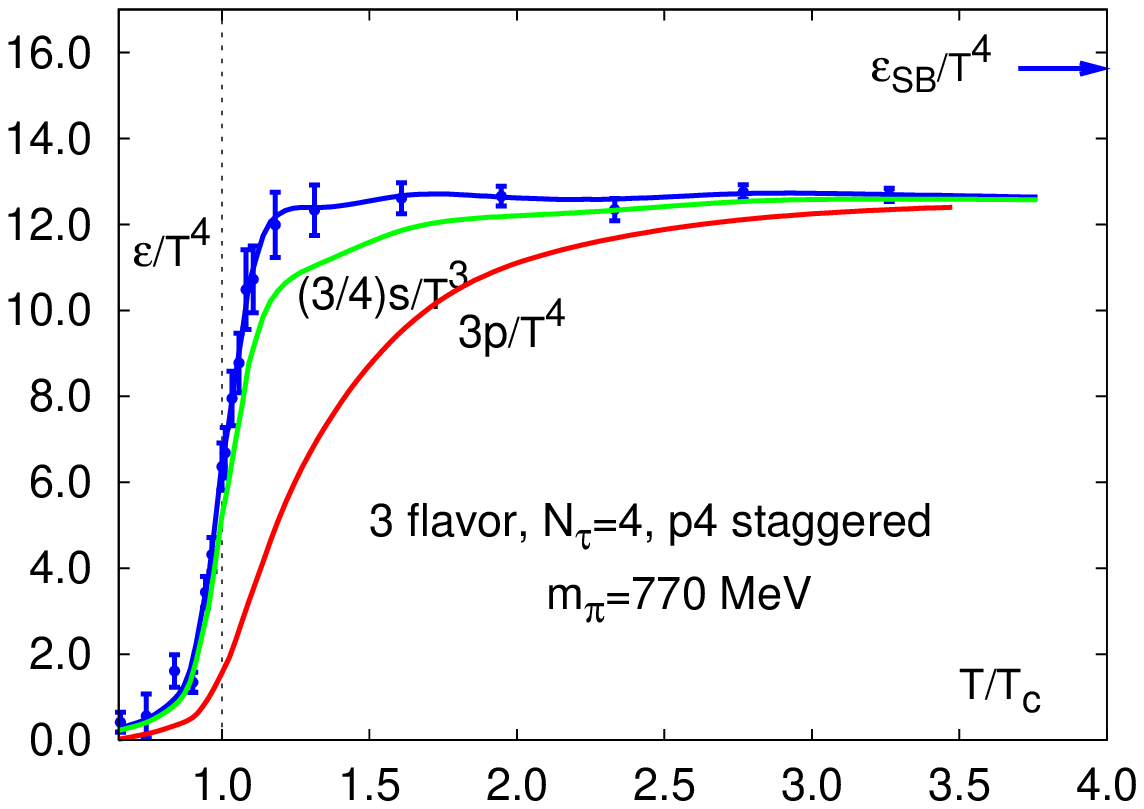,height=25pc}}
\caption{Figure of $\varepsilon(T)/T^4$, $P(T)/T^4$, and $s(T)/T^3$ for 
three light flavors of quarks on the lattice.}
\label{epsilon_vs_T}
\end{figure}

These computational advances have allowed us to glean several insights 
into the properties of hadronic matter at high temperature and vanishing 
net baryon density. 
The value of the transition temperature $T_c$ has been shown to lie 
in the range 160 -- 190 MeV \cite{Karsch:2001vs,Katz:2005br}
. For the physical values of the $u$, $d$ and
$s$ quark masses, the nature of the transition has been found to be 
a continuous cross-over, characterized by narrow, but finite peaks in various
susceptibilities \cite{Brown:1990ev}. 
The effective number of thermodynamically active 
degrees of freedom, defined as the ratio between the calculated 
entropy density and entropy density of a massless Bose field (SB):
\begin{equation}
\nu(T) = \frac{s(T)}{s_{\rm SB}(T)} \qquad {\rm with} \qquad
s_{\rm SB}(T) = \frac{2\pi^2}{45}T^3 ,
\label{eq1a2}
\end{equation}
rises very rapidly over a small temperature range around $T_c$ and
then levels off at about 80\% of the value $\nu_0=16+\frac{21}{2}N_f$ 
predicted for a thermal gas of non-interacting gluons and $N_f$ flavors 
of massless quarks, as shown in Figure~\ref{epsilon_vs_T}.

Many static properties of the high-temperature phase of QCD have 
been studied in detail by means of lattice simulations
\cite{Bernard:2004je}. Among 
these are the equation of state, i.~e.~energy density $\varepsilon(T)$
and pressure $P(T)$, quantities related to color confinement and 
spontaneous chiral symmetry breaking, such as the expectation value
of the Polyakov loop $\langle L\rangle$ and the quark condensate 
$\langle\bar\psi\psi\rangle$, various susceptibilities, and the 
potential $V_{\bar{Q}Q}(r)$ between a pair of heavy quarks
\cite{Kaczmarek:2002mc,Kaczmarek:2004gv}. The
results of these studies confirm that a dramatic rearrangement of
the internal structure of hadronic matter occurs near $T_c$, which
is consistent with the notion that states with open color are 
liberated above $T_c$ and chiral symmetry is restored, except for
the small current masses of the light quarks, which is not of 
QCD origin.

Much less is known with certainty about the dynamical properties 
of this quark-gluon plasma, because the Euclidean lattice gauge 
theory does not permit a direct calculation of quantities related
to real-time evolution. Some progress has been made toward the
calculation of dynamical response functions by analytic continuation
from imaginary to real time, but only in quenched QCD, i.~e.~in the
absence of dynamical light quarks \cite{Asakawa:2000tr}. Intriguing 
results of such calculations, indicating the survival of heavy-quark 
bound states up to temperatures of $2T_c$ or more, may well be 
artifacts of this uncontrolled approximation. Many dynamical 
properties of the high-temperature phase have been studied in the
framework of the so-called hard thermal loop (HTL) effective theory 
\cite{Braaten:1989mz}. There are good arguments in support of the 
notion that these techniques work reliably for temperatures $T>3T_c$ 
\cite{Andersen:2002ey,Blaizot:2003iq}, but it is not known precisely 
how low in temperature they can be applied and to which extent they 
work in the range of temperatures $T_c < T < 2T_c$ accessible in 
experiments at RHIC.

There have been speculations that the matter, which can
be produced and studied at RHIC, is very different from a plasma 
of weakly interacting quarks and gluons \cite{Shuryak:2004tx}. 
For common electromagnetic plasmas the well established measure of 
the interaction strength is the ratio between the average potential 
energy between neighboring ions and their kinetic energy: 
$\Gamma = E_{\rm pot}/E_{\rm kin}$. Plasmas with $\Gamma\ll 1$ are 
called weakly coupled, those with $\Gamma\gg 1$, strongly coupled
\cite{Ichimaru:2004}. Unfortunately, this criterion is
not applicable to matter composed of (nearly) massless, relativistic 
particles, such as the quark-gluon plasma, because the potential energy 
is not a well defined concept for such a system. A well defined measure 
of the interaction strength in a quark-gluon plasma is the ratio between 
its shear viscosity $\eta$ (crudely speaking a measure of the mean free 
path of particles) and its entropy density (a measure of the inter-particle 
distance) \cite{Kovtun:2004de}. In the weak coupling limit this ratio 
behaves like \cite{Baym:1990uj,Arnold:2000dr}
\begin{equation}
\eta / s \sim (\alpha_s^2 \ln 1/\alpha_s)^{-1} .
\label{eq1a3}
\end{equation}
A small $\eta/s$ ratio thus implies strong coupling. Since quantum 
mechanics limits the size of cross sections via unitarity, the 
ratio cannot be arbitrarily small. It has been conjectured that 
$\eta/s= 1/4\pi$ represents an absolute lower bound \cite{Kovtun:2004de}, 
and it has been shown that this value is reached in the strong coupling 
limit of certain super-symmetric gauge theories \cite{Policastro:2001yc}. 
Because lattice QCD is presently unable to reliably calculate transport
coefficients like the shear viscosity, we do not know what the value of 
$\eta/s$ is for QCD near $T_c$. 

As we shall discuss, experimental results from RHIC indicate that
the matter produced in nuclear reactions has a small ratio $\eta/s$.
This finding has reinforced speculations that the quark-gluon plasma near 
$T_c$ a strongly coupled one with the properties of a liquid with very 
low viscosity rather than a dilute gas \cite{Gyulassy:2004zy}. 
Whether this is connected with 
an internal structure that is dominated by quark-gluon bound states, 
as has been hypothesized \cite{Shuryak:2004tx}, remains unclear. Arguments 
have been given that, if a quasi-particle picture applies to the matter 
near $T_c$, these quasi-particles must carry the same quantum numbers as 
quarks \cite{Koch:2005vg,Ejiri:2005wq}. These arguments do not rule out 
a more complex structure of the gluonic component of the matter. 

It is appropriate to ask whether the quark-gluon plasma near $T_c$ 
displays other liquid-like properties besides a small viscosity. 
One property that often differentiates liquid forms of matter from 
gaseous ones is that liquids are generally rather incompressible, 
at least when compared with gases. Unfortunately, it is not clear 
how to apply this criterion to the quark-gluon plasma, since at 
net baryon number zero ($\mu_{B}=0$) and fixed temperature the 
pressure $P$ is independent of the volume $V$. Whether other aspects
of liquid-like behavior apply to the quark-gluon plasma near $T_c$
is an open question.

%% file: section1b.tex
The experimental observation and study of the quark-gluon plasma must 
be based on signals which can provide evidence of its formation and 
permit a characterization of its properties. Our evolving theoretical 
insight into the nature of this new state has also led to refinements 
in our understanding of these signals. Even if the transition between
matter composed of hadrons and the quark-gluon plasma state is only a 
rapid, but continuous cross-over, we can still make a clear observation 
of the new state if we understand some of the key properties that define 
it.

The experimental quest must proceed in steps: First, the determination
that the particles produced in the nuclear reaction really form, for a 
brief period, matter that deserves a description in thermodynamic terms; 
second, the determination that this matter has a novel structure and is 
not just a dense gas of hadrons; and third, the characterization of its 
main physical properties. In practice, these three steps are intertwined. 
In this section, we first give an update on some of the signatures 
discussed in the review by Harris and M\"uller \cite{Harris:1996zx} 
and then focus on the new probes acceessible at RHIC for the first time 
in heavy ion collisions, involving hard QCD scattering processes.

\subsubsection{Update on QGP signatures}

We start with an update of quark-gluon plasma thermodynamics and probes 
of physics at the thermal energy scale.

\begin{itemize}

\item
{\em Degrees of Freedom:} The lattice calculations predict that the
effective number of (massless) degrees of freedom $\nu$, as defined in 
Equation (\ref{eq1a2}), assumes a value around 30 nearly independent of 
temperature above $1.2\,T_c$. 
This behavior is very different from that of a hadronic 
gas, for which the effective number of degrees of freedom will steadily 
increase with temperature as more and more resonances are excited. It is 
important to note that the thermodynamic definition, Equation~(\ref{eq1a2}), 
of the number of degrees of freedom does not necessarily imply that these are 
associated with well-defined, particle-like excitations, usually called
quasiparticles. For temperatures near $T_c$, the quark-gluon plasma is 
quite likely strongly coupled, and we presently do not know what its 
dynamical excitations are. In fact, we do not know whether it contains any
well defined quasi-particles which account for the value of $\nu$ deduced 
from the lattice simulations. This is important, because it warns us that
an experimental determination of $\nu$ does not, in and of itself, tell us
that the elementary carriers of color, quarks and gluons, are deconfined
in the sense that they can propagate freely over distances larger than the 
size of a hadron (1 fm/c) without undergoing violent interactions.

\item
{\em Equation of State:} The lattice simulations of QCD tell us that the
speed of sound $c_s$, defined by $c_s^2 = \partial P/\partial\varepsilon$,
drops strongly below the value $c_s=c/\sqrt{3}$ valid for an ideal gas of
massless particles, as $T$ approaches $T_c$ from above ($c$ denotes the 
speed of light) \cite{Boyd:1996bx}. $c_s$ is expected 
to reach a minimum near $T_c$ and then increase again in the hadronic gas
domain. This has potentially observable consequences for the transverse 
expansion of the fireball. The minimum of the function $P/\varepsilon$ is 
called the ``softest point'' of the equation of state. If the matter is
produced with initial conditions near this point, the driving force of 
its expansion is minimal, leading to an increase in the lifetime of the 
fireball \cite{Hung:1994eq,Rischke:1996em}.

\item
{\em Fluctuations:} Statistical fluctuations of conserved quantum numbers 
in localized phase-space domains can be related to the corresponding 
susceptibilities of the matter. Susceptibilities measure the effect 
of a change in a certain intensive thermodynamic quantity, such as the 
temperature or a chemical potential on the associated extensive quantity,
such as entropy or a conserved quantum number. They thus tell us how the
system reacts to a change of external conditions. A second-order phase 
transition is characterized by certain divergent susceptibilities, such 
as the specific heat. Lattice simulations can make model independent 
predictions for such static quantities. From the experimental point of view, 
the most interesting quantities are the susceptibilities associated with 
the electric charge $Q$, strangeness $S$, and baryon number $B$
\cite{Asakawa:2000wh,Jeon:2000wg}. The lattice calculations predict that 
the associated susceptibilities change strongly across $T_c$. 
In particular, the flavor off-diagonal susceptibilities rapidly 
vanish above $T_c$ \cite{Allton:2005gk,Gavai:2005yk}, indicating that 
the thermodynamically relevant degrees of freedom carrying flavor and
electric charge have the precise quantum numbers of quarks and antiquarks, 
but not of quark-antiquark pairs as in a hadron gas \cite{Koch:2005vg}.

\item
{\em Chiral Symmetry Restoration:} The approximate chiral symmetry of QCD 
is spontaneously broken by the existence of a quark condensate in the 
vacuum. Lattice simulations predict a very rapid drop of the scalar quark 
condensate $\langle\bar{q}q\rangle$ from its vacuum value to almost zero 
in a narrow temperature region around $T_c$. The experimental observation 
of the predicted partial restoration of chiral symmetry above $T_c$ is more 
difficult, because quarks are predicted to retain sizable self-energies at 
high temperature \cite{Klimov:1981ka,Weldon:1982bn}. Although these do not 
break chiral symmetry, they are not easily distinguished from true scalar 
masses. 

One way of identifying the onset of chiral symmetry restoration would be 
an observation of the gradual disappearance of the mass splitting between 
hadronic states of opposite parity, but otherwise equal quantum numbers. 
The most promising way of achieving this is to measure the in-medium mass 
shifts of vector mesons and their chiral partners, such as the $\rho$ and 
$a_1$ mesons \cite{Rapp:1999ej}. 
This goal is complicated by the fact that the decays of the 
$a_1$ meson involve hadrons in the final state and thus are difficult to 
observe when they occur inside a dense medium.

Finally, a temporary restoration of chiral symmetry could manifest itself 
in the formation of local disorientated domains of the chiral order parameter
matrix $\langle\bar{q}_{\alpha}q_{\beta}\rangle$ ($\alpha$ and $\beta$ denote 
quark flavors). Such domains of disorientated chiral condensate (DCC) could 
be formed when the symmetry breaking is restored as the fireball cools 
rapidly through a second-order phase transition \cite{Rajagopal:1993ah} 
(for a review, see \cite{Mohanty:2005mv}). A related, but even more 
speculative phenomenon would be the formation of parity- or CP-violating 
metastable vacua of the pion field \cite{Kharzeev:1999cz}. 

\item
{\em Screening of Color Charges:} Perturbation theory predicts that the
forces between color charges are screened at high temperature. Recent 
lattice simulations have confirmed this prediction, but also generated 
some uncertainty with respect to the phenomenological consequences of 
color screening. Lattice determinations of the potential between a heavy
quark pair in a color singlet state \cite{Kaczmarek:2002mc}, combined with 
a potential model of heavy quarkonia, suggest that the charmonium ground 
state $J/\psi$ should cease to exist above $T_{\psi}\approx 1.1\,T_c$ 
and the ground state of the $(b\bar{b})$ system above 
$T_{\Upsilon}\approx 2.3\, T_c$ \cite{Digal:2001bh}. On the 
other hand, direct analysis of the spectral function of the $(c\bar{c})$
system in quenched QCD by means of the analytic continuation of imaginary
time correlators indicates the survival of a narrow $J/\psi$ state up to
at least $1.5\,T_c$ \cite{Asakawa:2003re,Datta:2003ww}. 
It is presently unclear whether this result is an artifact
of the quenched approximation or the analytic continuation procedure, 
or whether it indicates a failure of the potential model for charmonium
at finite temperature \cite{Mocsy:2005qw}.

Even if charmonium states survive significantly above $T_c$ as bound 
states or narrow resonances, their production in heavy ion collisions
may be strongly suppressed by reactions in the hot medium.  The 
($c\bar{c}$) pair, which can be produced in either the color singlet 
or color octet state, originally forms a highly localized state which 
then expands to hadronic size on a time scale dictated by the radius 
of the bound state (about 0.3 fm/c for $J/\psi$).  In the presence of 
a thermal medium the formation of the bound state may be jeopardized 
if the relative momentum of the pair is increased by elastic collisions 
with constituents of the medium during this formation time \cite{Qiu:1998rz}. 
Even after formation of the bound state, the pair can be dissociated by 
inelastic collisions, which excite the internal wave function above the 
open charm threshold \cite{Grandchamp:2003uw}.  

\item
{\em Thermalization:}
An important component of the phenomenology of probes of the hot medium
created in nuclear collisions at RHIC is the theory of initial formation 
of a high energy density state, its equilibration, and its evolution.
Much progress on understanding these problems has been made over the 
past decade. A semiclassical description of the initial-state gluon 
fields in the colliding nuclei has
allowed theorists to make detailed predictions for the space-time and
momentum space distribution of partons in the pre-equilibrium phase of 
the reaction \cite{Blaizot:1987nc,Kovner:1995ja,Kovchegov:2000hz}. It 
is now believed that perturbative QCD processes do not lead to a very
rapid equilibration and thermalization of the initial parton distribution
\cite{Biro:1993qt,Baier:2000sb,Molnar:2001ux} 
(but see \cite{Wong:1996va,Xu:2004mz} for contrasting opinions).  
An alternative pathway to thermalization, which has recently been studied 
extensively is furnished by instabilities among the soft gluon modes in a 
partonic medium with an anisotropic momentum distribution 
\cite{Mrowczynski:1994xv,Arnold:2004ti,Rebhan:2005re,Dumitru:2005gp,Mrowczynski:2005ki}.
It is not yet clear whether this mechanism reduces the thermalization
time significantly compared to the perturbative thermalization scenario
or just causes the parton distribution to rapidly become isotropic 
\cite{Arnold:2003rq,Mueller:2005un}.

\item
{\em Dynamical Properties:} Lattice simulations are presently unable to
make reliable predictions of most dynamical properties of the quark-gluon
plasma. Recent attempts to determine the real-time spectral functions for 
excitations of various color-singlet channels 
\cite{Asakawa:2000tr,Karsch:2001uw,Petreczky:2003iz} are limited by the 
need to use the quenched approximation, and the results show inconsistencies, 
especially for low frequencies \cite{Blaizot:2005mj}. 
The calculation of phenomenologically relevant transport properties, 
such as the shear viscosity or collective modes, remains an important 
challenge \cite{Petreczky:2005zy}.

However, recently there has been important progress in calculating 
these dynamical properties perturbatively in a dual quantum field theory 
involving black holes in anti-de Sitter (AdS) space \cite{Kovtun:2004de}.  
This approach is based on the insight derived from string theory that 
weakly coupled gravity theories in higher dimensions can be dual to 
four-dimensional gauge 
theories in the strong coupling limit \cite{Maldacena:1997re}.  It must
be emphasized that these AdS/CFT (conformal field theory) techniques 
presently have the limitation that no higher dimensional gravity or 
string theory is known which is dual to QCD.  However, recent progess
in calculating hadron structure and scattering processes make this an 
area to watch in the future~\cite{adscft_progress}.

\end{itemize}

\subsubsection{New probes of the medium}

One of the greatest recent advances in heavy ion physics is that, 
at the RHIC energies, one finally has access to hard processes where 
perturbative QCD (pQCD) calculations provide a solid theoretical
framework \cite{Satz:1995cg}.  Ideally we would like to probe the created 
medium at different times as the system is expanding and becoming more 
dilute.  One limitation is that all probes of the medium must be generated 
in the collision itself, and we must quantitatively understand their initial 
production cross sections in order to consider them as controlled and 
calibrated probes. The following list covers a few of the hard probes of 
dense matter relevant to the experiments at RHIC.  

\begin{itemize}

\item
{\em Factorization of QCD cross sections:}
Parton-parton reactions involving large momentum transfer can result 
in back-to-back jets which have been well described at hadron colliders 
in terms of 
pQCD parton-parton cross sections augmented by experimentally determined
initial parton distribution functions (PDF) and final fragmentation 
functions (FF).  This framework assumes that the hard scale of the 
parton-parton interaction can be separated (``factorized'') from the soft, 
non-perturbative physics governing the PDF and FF. It further assumes 
that these functions are universal, i.e. they can be determined from well
understood processes, like deep-inelastic scattering or 
$e^{+}e^{-} \rightarrow$~jets.  If this factorization and universality 
can be shown to apply at the modest $p_{T}$ range accessible at RHIC 
($p_{T}^2\approx 10-1000$~GeV$^2/c^2$), then perturbative quarks and 
gluons constitute a calibrated probe which can be used to determine 
properties of the medium, for example, the color opacity 
\cite{Accardi:2003gp}. We will show in Section \ref{sect2c} that the
RHIC data, indeed, support this notion.

\item
{\em Nuclear modifications of parton distributions:}
One caveat is that it is well established that PDF are modified in nuclei 
relative to the proton by processes called shadowing, EMC effect, etc.  
For recent reviews see \cite{Accardi:2003be,Norton:2003cb}.  
Partons carrying a very small momentum fraction $x$ of the parent nucleon 
are generally depleted in nuclei compared with the proton.  One model for 
understanding this depletion of low-$x$ partons in nuclei asserts that the 
rapid growth of gluons must saturate when the density reaches occupation 
numbers of order $\alpha_s^{-1}$ \cite{Gribov:1984tu}.
In this saturation scenario the nucleus is more appropriately described 
as a single gluonic wave function, sometimes referred to as the Color 
Glass Condensate (CGC) \cite{McLerran:2002wj}. Because the scale $Q_s$ 
at which gluon saturation occurs grows with nuclear mass 
\cite{Mueller:1985wy}, it has been proposed that this process may be 
described perturbatively for large nuclei, especially at very high energies
\cite{McLerran:1993ni}.  A key test of the importance of nuclear 
modifications of parton distributions already at RHIC energies 
was to check if one could observe significant suppression effects 
on jets as an initial state effect from the nuclear ($^{197}$Au) 
wave function \cite{Kharzeev:2002pc}.

\item
{\em Parton energy loss:}
Once an energetic parton is injected into the medium via a hard 
scattering event, it can lose energy and momentum by rescattering 
within the medium. We can distinguish three types of interactions of 
the parton with the medium: If the parton elastically scatters on a 
structureless color charge, this will not result in an appreciable 
energy loss but in a random walk of transverse kicks.  In order to scatter 
inelastically, a parton can either transfer some of its energy to a 
bound constituent of the medium or lose energy by radiating a gluon. 
In all three cases, the interaction strength (and in the latter two, 
the energy loss) is proportional to the density of color charge carriers 
in the medium and square of the color charge of the parton. Thus, a 
gluon, whose color charge is $3/\sqrt{2}$ times as large as the color
charge of a quark or antiquark, should lose approximately twice as much 
energy as a quark in the same medium. 

For light quarks and gluons with a large momentum, the dominant 
mechanism of energy loss is expected to be gluon radiation. 
The radiative energy loss $\Delta E$ of a massless parton penetrating 
a strongly interacting medium can be expressed in terms of a transport
coefficient $\hat{q}$ characterizing the stopping power of the medium
and defined as \cite{Baier:1996kr,Baier:2000mf}
\begin{equation}
\hat{q} = \rho \int dq^2 q^2 \frac{d\sigma}{dq^2} 
= \langle q^2 \rangle / \lambda_f ,
\label{eq3b1}
\end{equation}
where $\rho$ denotes the density of scattering centers, $\sigma$ is the
elastic scattering cross section, and $\vec{q}$ the momentum transfer 
in a collision of the parton with a medium constituent. $\lambda_f$ is
the mean free path of the parton in the medium. In a static, homogeneous 
medium, the average energy loss depends quadratically on the penetration 
length $L$. The unusual dependence results from the rescattering
of the radiated gluon in the medium in combination with the coherence
of the radiation from scatterings occurring in rapid succession
(called the Landau-Pomeranchuk-Migdal effect 
\cite{Landau:1953um,Migdal:1956tc}). 

For a comparison with the RHIC data, one needs to take into account that 
the medium created in a nuclear collision at RHIC is not static and 
homogeneous. This can be done by defining an effective path length
\cite{Salgado:2003gb}.
In addition, it is important to use the full energy loss distribution 
$P(\Delta E)$ instead of the average energy loss $\overline{\Delta E}$ 
\cite{Baier:2001yt}.  It is also important to recognize that the 
requirement of emission of an energetic hadron increases the weighting 
toward large values of the fragmentation variable $z$ and favors the 
near-surface emission of the hard parton 
\cite{Shuryak:2001me,Muller:2002fa}. 

\item
Because their creation represents a hard QCD process, which can be 
calculated within the framework of pQCD, heavy quarks can also serve 
as hard probes of the hot medium \cite{Bedjidian:2003gd}. Besides 
their role as probes of color screening, heavy quarks 
also probe the color charge density in the medium through their 
energy loss, when they are created with a sizable transverse momentum.
Their interaction with the medium differs from that of light quarks 
and gluons, because at leading order in the QCD coupling constant
heavy quarks only interact with gluons in the medium. In addition, 
gluon radiation from 
a heavy quark with mass $M$ and energy $E$ is strongly suppressed at 
forward angles $\theta < \theta_0 = M/E$ \cite{Dokshitzer:2001zm}.  
This ``dead cone effect'' \cite{Dokshitzer:1991fd} reduces the radiative 
energy loss of heavy quarks for moderate $p_{T}$. 

\end{itemize}

%% file: section1c.tex

One motivation for the construction of RHIC was the desire to explore
the physics of matter under conditions that existed during the first 
$10-20~\mu$s of our universe in controlled experiments.\footnote{In 
the inflationary Big Bang model, the universe first inflates in an
ultracold (false vacuum) state and then reheats to a high temperature
after making the transition into the true vacuum. Usually one assumes 
that the reheating temperature is far above the electroweak transition
($T \gg 100$~GeV). But the observational evidence only demands that 
the temperature generated by reheating after inflation exceeds about
4 MeV \cite{Hannestad:2004px}. It is thus not completely certain that
the temperature range reached in the RHIC experiments was ever probed
in the Big Bang.} If the deconfinement transition were a strong first
order phase transition, it would have proceeded via bubble nucleation 
\cite{DeGrand:1984uq}. The formation of macroscopic bubbles would have 
led to violent processes that would have left traces observable today 
(for a recent review see \cite{Schwarz:2003du}). The two most important 
processes associated with bubble formation
are the creation of local inhomogeneities in the cosmic baryon density 
during the time of nucleosynthesis and the possible formation of 
(meta-)stable strange quark matter nuggets \cite{Witten:1984rs}. 
Both of these possibilities appear to be ruled out by the absence of
a strong first-order phase transition in the equation of state predicted
by lattice gauge theory. 

One observable effect of the cosmological QCD phase transition, in 
principle, is the modulation of the spectral density of primordial
gravitational waves due to the modification of the Hubble expansion 
as a result of the change in the number of light degrees of freedom 
\cite{Schwarz:1997gv}. The modulation caused by the QCD transition 
is almost independent of the type of transition. If strongly first
order, the transition itself could also have generated gravitational
waves due to collisions among large nucleating bubbles \cite{Witten:1984rs}. 
The interesting frequency range of gravitational waves is near 100 nHz. 
Unfortunately, no technique capable of observing gravitational waves 
in this frequency range is known today.  Thus, the study of heavy ion 
collisions may provide the best insight on this early epoch in our universe, 
rather than the observation of the cosmological transition itself.

The unanticipated liquid-like nature of the quark-gluon plasma 
created in collisions at RHIC suggests analogies to the physics of
strongly coupled electromagnetic plasmas \cite{Thoma:2005aw}.
For such plasmas, the viscosity is large both in the limits of weak 
and strong coupling and attains a minimum at intermediate coupling.
Interestingly, the naive application of classical measures of the
coupling strength to the quark-gluon plasma near $T_c$ also gives 
values in this intermediate range of values of the interaction 
measure \cite{Thoma:2004sp}.

Phenomena analogous to the anisotropic collective expansion of the
matter produced in heavy ion collisions have recently been observed
in degenerate systems of fermionic atoms released from optical traps
\cite{li6ref}. These systems exhibit many features 
associated with superfluidity, and their collective flow is probably 
governed by collisionless ideal fluid dynamics. It is an intriguing
question whether the physics underlying the nearly inviscid transverse
flow of matter produced in relativistic heavy ion collisions is of a 
similar nature.

%% file: section1d.tex
The Relativistic Heavy Ion Collider construction was completed in 1999 with the
first physics running the following year.  
The collider is located in a 3.8 km circumference tunnel and is composed of
two identical, quasi-circular rings of superconducting magnets oriented
to intersect at six experimental hall locations~\cite{rhic_review}.  
The machine can accelerate gold nuclei up to momenta of 100 GeV/c per
nucleon in each beam and protons up to 250 GeV/c, and smaller mass nuclei
up to intermediate momenta depending on their mass to charge ratio.
The center-of-mass energy in these collisions is over an order of magnitude 
higher than the previous highest energy heavy ion reactions at the CERN 
SPS fixed target facility, and two orders of magnitude above the BNL AGS 
fixed target facility. 
The nominal design luminosity for gold nuclei at full energy of 
$2 \times 10^{26}~cm^{-2}s^{-1}$ was already achieved during the first two physics 
running periods.

There are four dedicated experiments for studying 
heavy ion physics at RHIC:  BRAHMS, PHENIX, PHOBOS and STAR. Detailed 
descriptions of the detectors and their performance are given in 
\cite{rhic_nim_volume}.
The experimental program was designed such the experiments 
contain not only unique capabilities, but significant overlap 
to provide essential cross checks.  

The BRAHMS experiment has 
two movable, small acceptance spectrometer arms.  A range of detector 
technologies allow them to identify hadrons ($\pi, K, p$) up to very large 
rapidity when the forward arm is positioned close to the beam line.  
The PHOBOS experiment utilizes silicon technology and covers nearly the full 
solid angle for charged particles, but with limited 
particle identification.  Thus, the PHOBOS experiment has excellent global 
event characterization capability.  

\begin{figure}
\centerline{\psfig{figure=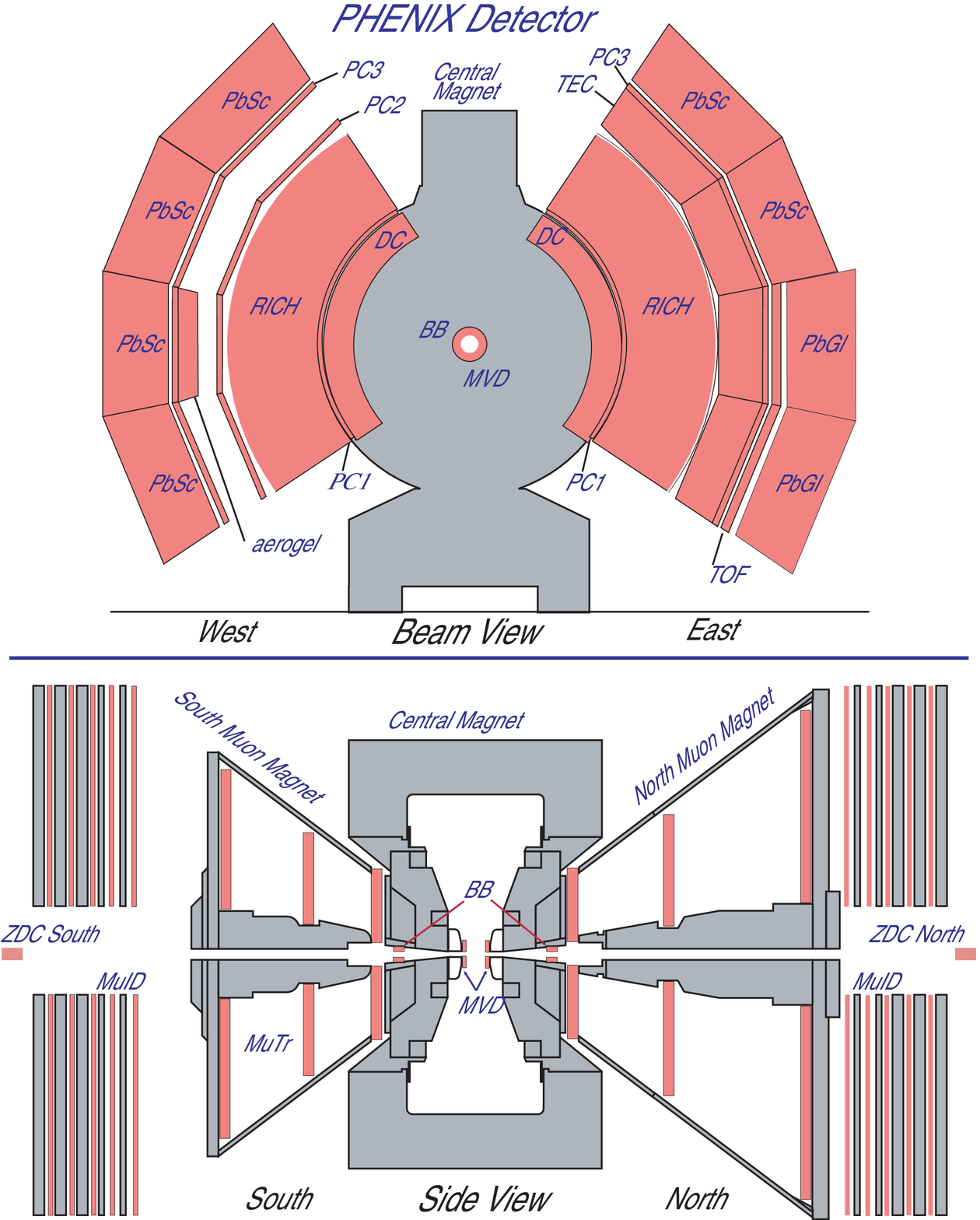,height=25pc}}
\centerline{\psfig{figure=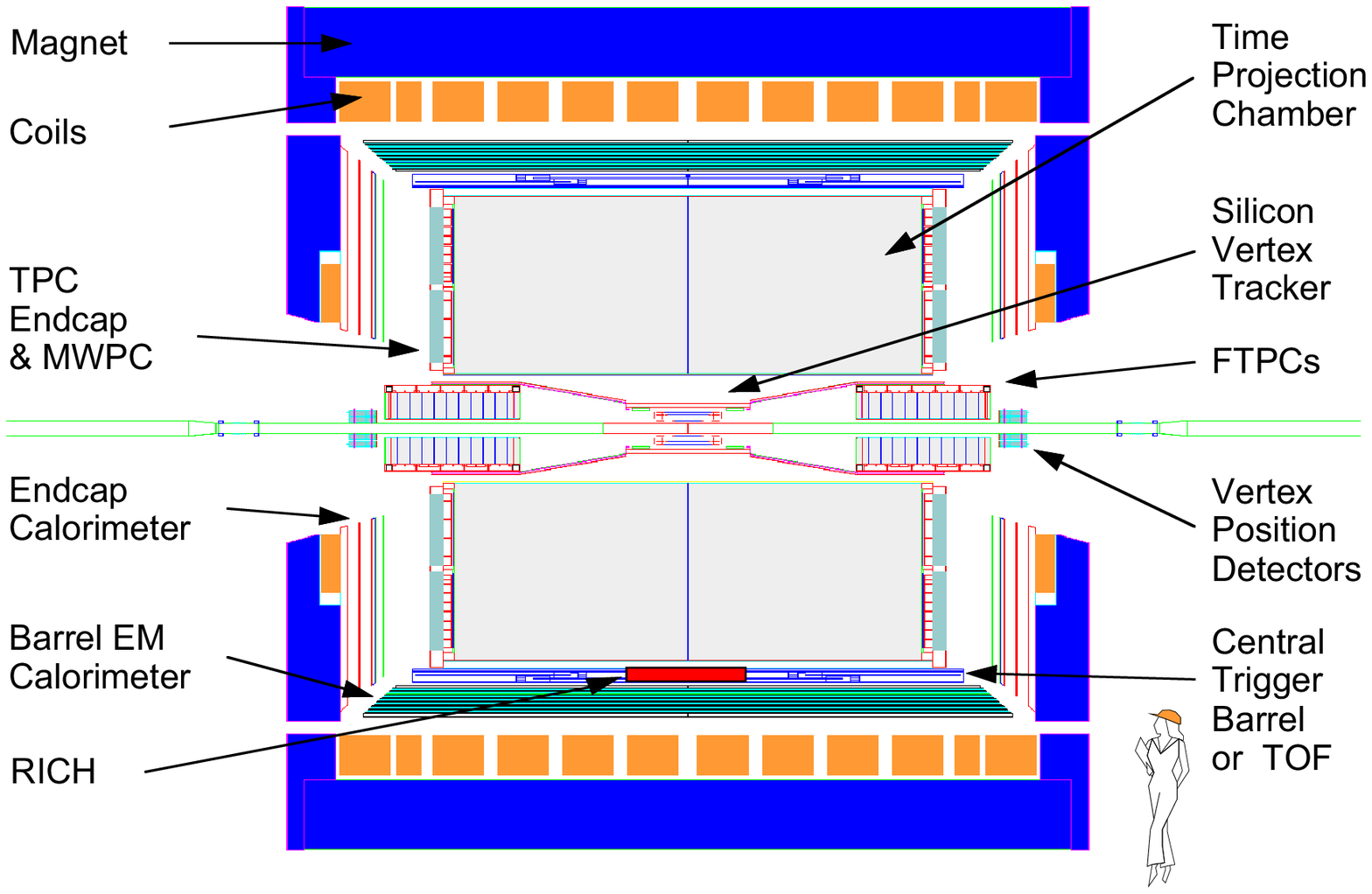,height=25pc}}
\caption{Schematic diagrams of the PHENIX (above) and STAR (below) experiment configurations.  Details of the
various detector subsystems are given in~\cite{rhic_nim_volume}}
\label{fig_detect}
\end{figure}

The PHENIX experiment is a combination of two detector systems.  
There are two spectrometers with limited rapidity coverage around 
midrapidity with excellent hadron, electron and photon identification 
capabilities.  Additionally there are two muon spectrometers at 
forward and backward 
rapidity for measuring decays of heavy quarkonia states and single muons.  
A high speed trigger and data acquisition allow for nearly dead-timeless
running.
The STAR experiment is based around a large coverage Time Projection Chamber 
(TPC) inside a solenoidal magnet 
that provides exceptional charged particle tracking and  
particle identification via the measurement of ionization energy loss $dE/dx$.  
The large coverage allows for multi-particle
correlation measurements and also reconstruction of multi-strange baryon 
decays.  In addition, STAR has an electromagnetic calorimeter and additional
forward angle detectors including a radial-drift TPC.  
The STAR data acquisition is high bandwidth and incorporates a
multi-tiered trigger system.
Schematic diagrams of the two large detectors are shown in Figure~\ref{fig_detect}.

All experiments have been successful in being quickly commissioned 
and providing interesting physics results from the first running at RHIC.  
The accelerator has the flexibility to run multiple species and energy 
configurations each year.  The different colliding species, energies and integrated 
luminosities to date are shown in Table~\ref{table_rhic}.

\begin{table}%
\def~{\hphantom{0}}
\caption{Table of RHIC Performance.}\label{table_rhic}
\begin{tabular}{@{}lcccc@{}}%
 
\toprule
Run  &  Species   &  Particle Energy &   Total Delivered & Average Store  \\
 &    &  $[GeV/n]$  &   Luminosity    & Polarization \\
\colrule
Run-1 2000   & $Au + Au$   & 27.9   &  $< 0.001 \mu b^{-1}$   & - \\
             & $Au + Au$   & 65.2   &  20 $\mu b^{-1}$        & - \\
Run-2 2001-2 & $Au + Au$   & 100.0  &  258 $\mu b^{-1}$       & - \\
             & $Au + Au$   & 9.8    &  0.4 $\mu b^{-1}$       & - \\
             & pol. $p+p$  & 100.0  &  1.4 $\mu b^{-1}$       & 14\% \\
Run-3 2002-3 & $d + Au$    & 100.0  & 1.4 $pb^{-1}$           & - \\
             & pol. $p+p$  & 100.0  & 5.5 $pb^{-1}$           & 34\% \\
Run-4 2003-4 & $Au+Au$     & 100.0  & 3740 $\mu b^{-1}$       & - \\
             & $Au+Au$     & 31.2   & 67 $\mu b^{-1}$         & - \\
             & pol. $p+p$  & 100.0  & 7.1 $pb^{-1}$           & 45\% \\
Run-5 2004-5 & $Cu+Cu$     & 100.0  & 42.1 $nb^{-1}$          & - \\
             & $Cu+Cu$     & 31.2   & 67 $\mu b^{-1}$         & - \\
             & $Cu+Cu$     & 11.2   & 0.02 $nb^{-1}$          & - \\
             & pol. $p+p$  & 100.0  & 29.5 $pb{-1}$           & 46\% \\
             & pol. $p+p$  & 204.9  & 0.1 $pb^{-1}$           & 30\% \\
\botrule
\end{tabular}
\end{table}

During the early discovery phase, one focus of the experiments at RHIC has been
establishing evidence for the formation of a thermodynamically equilibrated
state of matter in gold-gold collisions. In addition, they have begun to address
the following properties of the created medium:

\begin{itemize}
\item Its initial energy and entropy density;
\item Its flavor composition and net baryon content;
\item Its equation of state;
\item Its shear viscosity;
\item Possible collective modes;
\item The quantum numbers of the carriers of collective flow;
\item The characterization of jet quenching;
\item Heavy quarkonia suppression;
\end{itemize}

The experiments have published an extensive set of refereed publications to date.  In 
addition, each experimental collaboration has published a review of all their results
to date in a single volume~\cite{BRAHMS_Whitepaper,PHENIX_Whitepaper,PHOBOS_Whitepaper,STAR_Whitepaper}.
\footnote{In general we will reference the original data publication, except in cases where the
results are a compendium of results and we will refer to these so-called ``White papers.''}

%% file: section2a.tex
When colliding two nuclei one wants to understand how the initial kinetic 
energy (39 TeV at RHIC in gold-gold reactions) is re-distributed after the 
collision in terms of particle production, energy per particle, collective 
motion, etc.  An intimately related question is that of entropy production, 
since the initial nuclei form a very low entropy state and the final 
configuration has a very large entropy.  

The BRAHMS experiment measures the rapidity distribution of protons and 
thus determines how much kinetic energy appears to be retained by the 
colliding nucleons~\cite{brahms_stopping}.  Their data indicate that in 
central gold-gold reactions, over $28\pm 3$ TeV of energy is deposited 
in heating the newly created medium and collective motion.  All four experiments measure the resulting 
distribution of this energy amongst newly created particles.  The PHOBOS 
experiment with measurements over nearly the full solid angle observes the 
creation of approximately 5000 charged particles and thus 7500 total particles, 
based on estimates of the neutral particle contribution~\cite{phobos_mult}.  
The production of nearly 20 particles per incident nucleon implies a 
large entropy production.  A critical task is determining when and how 
this entropy is produced.

The experiments have also measured the charged particle multiplicity as
a function of collision energy and centrality (impact parameter).
The growth of multiplicity with increasing collision
energy~\cite{BRAHMS_Whitepaper,PHENIX_Whitepaper,PHOBOS_Whitepaper,STAR_Whitepaper} 
is less than expectations from in a simple picture of ``soft'' coherent production 
combined with ``hard'' mini-jet production that increases quickly with 
increasing energy~\cite{soft_hard}.  
An alternative description of the multipicity dependence was proposed
in the context of saturation models where the incoming gluon density at low momentum
fraction $x$ in the incoming nuclei is suppressed.  
Color Glass Condensate (CGC) models which incorporate gluon saturation
have had significant success in describing the data~\cite{cgc_mult}.  Detailed 
tests of whether this picture captures the underlying physics are discussed 
later in the context of hard processes (see Section~\ref{sect2c}).  

The PHOBOS experiment has made the observation that the charged particle multiplicity
per participating nucleon pair in central gold-gold reactions is nearly identical to that 
measured in $e^{+}e^{-}$ annihilation events at the same center-of-mass energy, again per
participant pair for the heavy ion case~\cite{PHOBOS_Whitepaper}.  The agreement persists over the
entire energy range at RHIC.  Such scaling observations are striking and have yet to 
result in a detailed explanation in terms of, for example, global energy conservation.  This
agreement is particularly puzzling since the $e^{+}e^{-}$ growth in multiplicity with energy is 
well described by perturbative QCD in terms of gluon splitting and then mapping partons into hadrons
via parton-hadron duality~\cite{mueller_ee}.

\begin{figure}
\centerline{\psfig{figure=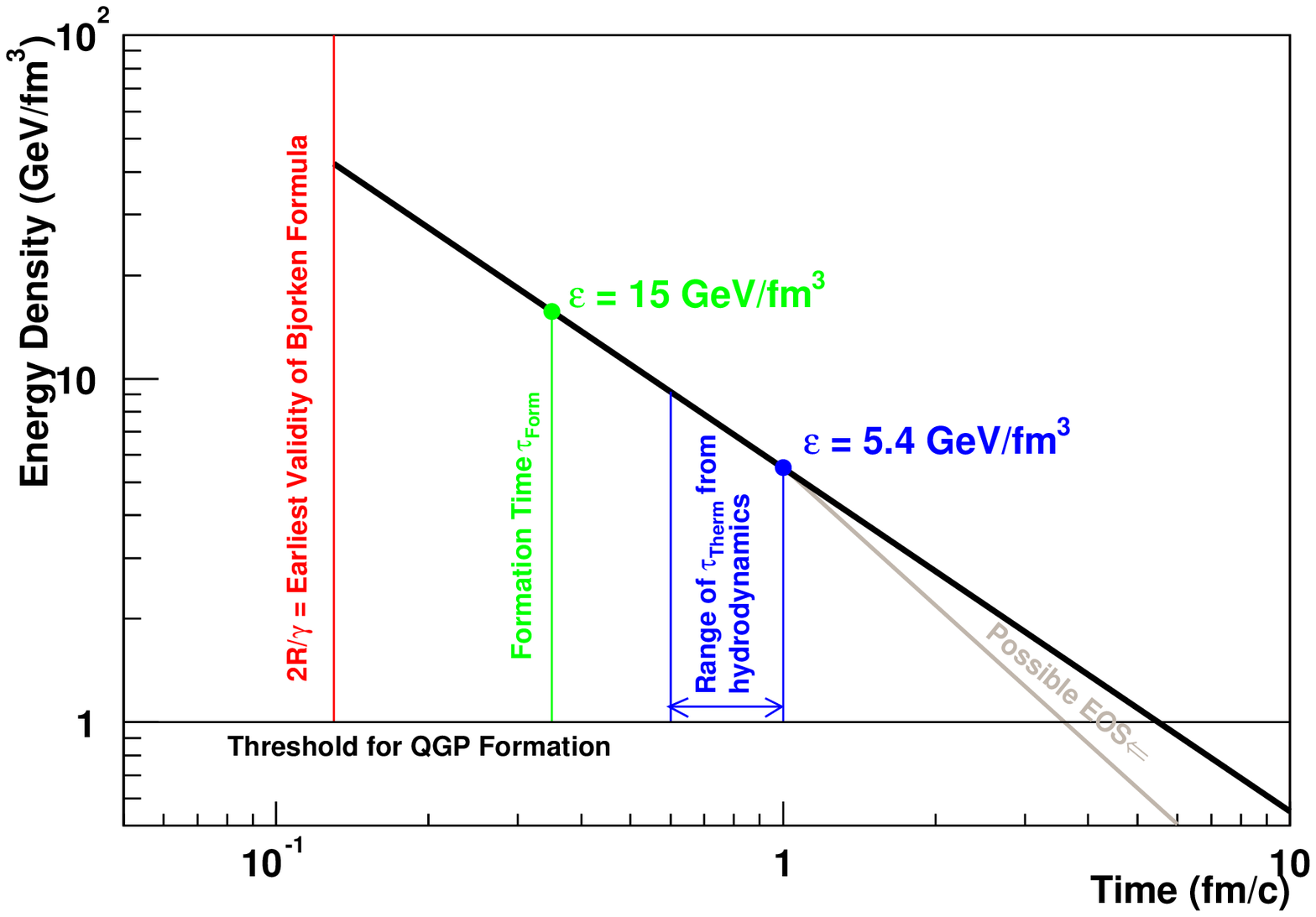,height=25pc}}
\caption{Schematic drawing of the time and energy density scales derived through the Bjorken picture.
Taken from~\cite{PHENIX_Whitepaper}.}
\label{fig_edense_time}
\end{figure}

Experiments have also measured the total transverse energy
production in these reactions.\footnote{The transverse energy measurements to date are centered 
at mid-rapidity.  Many of the experimental observations are in a window around mid-rapidity.  
We will make special note of important measurements at forward and backward rapidity.}
One can relate the transverse energy to the
energy density in the context of a boost-invariant cylinder scenario~\cite{bjorken_scenario}
which suggests an initial energy density of 15~GeV/fm$^{3}$ decreasing to
5.4~GeV/fm$^{3}$ by a time $t=1$~fm/c as shown in Figure~\ref{fig_edense_time}~\cite{PHENIX_Whitepaper}.
Although the energy density decreases as $1/$time or faster, the energy density 
significantly exceeds the expected transition density
from lattice QCD calculations for a time up to approximately 5~fm/c.
It is notable that the CGC calculations when tuning the saturation scale 
$Q_{S}$ to match the multiplicity, overpredict the transverse energy (or 
$E_{T}$ per gluon which translates using parton-hadron duality into $E_{T}$ 
per hadron).  This may be reconciled if there is substantial longitudinal work done
in the expansion which will decrease the total transverse energy.
Dynamical models describing the three-dimensional 
evolution, which have recently become available \cite{Hirano:2003pw}, allow
for the study of these effects in detail.

\begin{figure}
\centerline{\psfig{figure=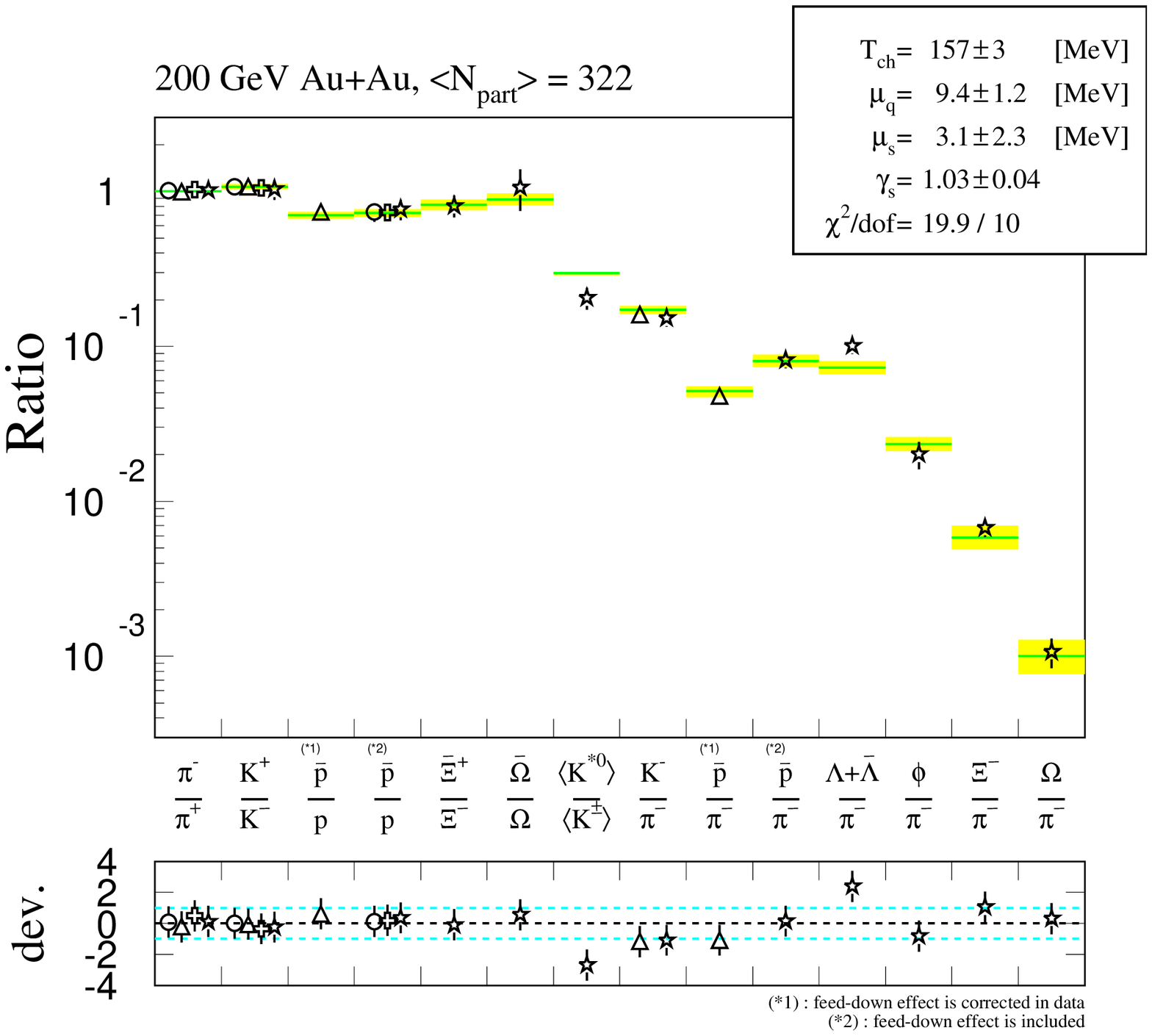,height=25pc}}
\caption{Comparison of BRAHMS (circles), PHENIX (triangles),
PHOBOS (crosses) and STAR (stars) particle ratios from central gold-gold
collisions at $\sqrt{s_{NN}}=200~$GeV at mid-rapidity.  The thermal
model descriptions from~\cite{kaneta} are also shown as lines.  Similar
results are obtained in~\cite{pbm}.}
\label{fig_ratios}
\end{figure}

Almost the entire particle production can be described with a few simple 
parameters.  One can calculate the yields of most hadrons 
($\pi, K, p,\overline{p}, \Lambda, \Sigma, \Xi, \Omega, \phi$) 
near midrapidity in terms of a single temperature of order 170 MeV and 
a small light quark (baryon) chemical potential ($\mu_{B}$), as shown in 
Figure~\ref{fig_ratios}~\cite{BRAHMS_Whitepaper,PHENIX_Whitepaper,PHOBOS_Whitepaper,STAR_Whitepaper}.
\footnote{Hadron abundances containing charm quarks are not 
described by chemical equilibration as expected since their quark mass significantly exceeds
the temperature.  Additionally, unstable resonances such as the $K^{*}$ deviate
from chemical equilibration~\cite{STAR_kstar} potentially due to re-scattering of their daughter decay products.
}
One particularly striking feature is that even the multi-strange particles 
appear at chemically equilibrated levels.

The transverse momentum distributions of the emitted particles have also 
been measured,
and in many cases the rapidity distributions as well~\cite{BRAHMS_Whitepaper,PHENIX_Whitepaper,PHOBOS_Whitepaper,STAR_Whitepaper}.
The mean transverse momentum $\langle p_T\rangle$ for pions is of order 
400 MeV. If we translate this into three dimensions and multiply by the 
particle number (7500), we obtain only about 4 TeV.  Thus, a very large 
fraction of the incident energy resides in the (collective) longitudinal 
motion of the medium.  The PHOBOS and BRAHMS experiments in particular 
have characterized the longitudinal motion in rapidity space~\cite{brahms_rapidity,PHOBOS_Whitepaper}.  
The identified pions follow a Gaussian distribution in rapidity with a width 
consistent with the prediction of the Landau hydrodynamical model~\cite{landau}.  
In this scenario, all of the incoming energy is rapidly thermalized in a 
volume the size of the Lorentz contracted overlapping nuclei which then 
expands hydrodynamically until the particles decouple.  This model would 
imply all the entropy generation takes place in the very short overlap 
time of the two nuclei, $2R/(\gamma c)<0.2$~fm/c, followed by isentropic
expansion.  It would imply an enormous initial equilibrated energy density of about
3000 GeV/fm$^3$.

In the alternative Bjorken scenario~\cite{bjorken_scenario} much of the 
initial incoming 
longitudinal momentum is transformed into longitudinal collective 
motion, and the system can be described as a boost-invariant cylinder of 
fireballs distributed in rapidity space.  
The Bjorken scenario is often invoked for simplicity, but the measured
rapidity distributions do not show boost invariance.  
In addition, particle ratios and spectra are observed to vary significantly as 
a function of rapidity~\cite{ratios_rapidity}. The deduced baryon chemical 
potential is larger at forward and backward rapidities, as one might 
expect from the increased presence of remnants of the colliding nuclei 
in these regions of phase space.   
This observation shows that the system is intermediate between the
limits of the Bjorken and Landau models.

The transverse momentum spectra of the particles are found to deviate from 
a static Fermi-Dirac or Bose-Einstein distribution (approximately Boltzmann 
for the temperatures and chemical potentials relevant for RHIC).  However, the spectra are well 
described at low $p_T<2$~GeV/c with a common Boltzmann component and a 
radial velocity boost from the outward blast of the 
medium~\cite{BRAHMS_Whitepaper,PHENIX_Whitepaper,PHOBOS_Whitepaper,STAR_Whitepaper}.  Simultaneous
constraints from the transverse momentum spectra of many different hadron 
species (with a wide range of masses) and two-particle correlation 
results constrain these contributions.  If one removes the common velocity 
boost component, one finds the measurements 
at midrapidity are consistent with black-body radiation from a hot medium. 
The particle ratios indicate that the emissivity of the fireball is nearly the
same (most likely unity) for all stable hadrons.  

It has been noted that the appearance of an exponential energy distribution 
of particles is not conclusive evidence for equilibration but can also be the result 
of phase space dominance of the final states.  
However, in elementary particle reactions, particles carrying non-zero 
strangeness are always suppressed~\cite{becattini}, whereas in the nuclear reactions at 
RHIC their yields are in perfect agreement with the predictions 
of the thermal model.  The simplest explanation of this observation is that
the available energy is equilibrated over a large enough volume to 
approximate a grand canonical ensemble, including strangeness.  In addition, there is significant 
evidence for intense re-scattering amongst constituents to favor a thermal
intermediate state over a direct reaction with a final state dominated by 
phase-space. 

One additional set of global observations relate to fluctuations.  In any 
study of phase transitions, the measurement of particle number and energy 
fluctuations is very relevant.  If the transition were strongly first order, 
one might expect the fireball evolution to involve large bubble formation 
from a supercooled phase, which could lead to large phase-space fluctuations 
within a single event and also significant event-by-event fluctuations.  
Predictions were also made that one could have domains (regions in phase 
space) with disorientated chirality (DCC, see Section~\ref{sect1b}) that would 
re-align their chirality through the emission of low $p_T$ pions with 
anomalous isospin distributions~\cite{Mohanty:2005mv}.  There have also been predictions of 
the formation of domains with strong CP violation~\cite{CP_bubbles}.  

%
%
These measurements are quite challenging and published results to date~\cite{STAR_fluct,PHENIX_Whitepaper}
already indicate that multiple contributions to the fluctuations are likely.
The non-statistical fluctuations may have contributions from 
jet fragmentation correlations, quantum statistical correlations,
and for the case of hadrons with complementary quantum numbers (baryon-antibaryon or positive-negative 
charge pairs) conservation correlations (as described by balance functions~\cite{balance_functions}).  
No unambiguous statement on fluctations from a particular phase transition can be made
at present.

%% file: section2b.tex
One of the most dramatic discoveries at RHIC is that the medium displays 
a high degree of collectivity, often referred to as flow.  Since we 
compress nuclear matter in the early stages of the collision, 
if there are large interaction cross 
sections between constituents of the medium, we expect density gradients 
to translate into outward pressure that cause the system to explode at 
relativistic speeds.  In the previous section in discussing the $p_T$ 
spectra of various hadrons, it was necessary to include an outward boost 
velocity  $v_{b}/c > 0.5$ to describe the mass dependent differences.  

Experiments can probe this outward pressure with great precision by 
characterizing non-central (intermediate impact parameter) nuclear 
collisions.  In these reactions, the 
nuclear overlap region is not circular in the transverse plane, but
instead elliptically shaped.  Thus, the density distribution if decomposed 
into azimuthal angle Fourier components has many non-zero coefficients, 
with the second coefficient being by far the largest. The 
stronger the interactions amongst the particles, the greater the 
translation of spatial anisotropy into momentum anisotropy of the final 
observed particles.  There are two key features of this flow behavior.  
First, the flow is self-limiting in that as the system expands the spatial 
distribution becomes more isotropic.  Thus, a large contribution to the 
flow must come from interactions in the first $2$~fm/c after the collision~\cite{heinz_flow}.  
Second, regardless of the constituents of the matter at these early times, 
the momentum anisotropy is preserved through any type of transition to the final 
observed hadron distributions.

Experimentally one measures the $\phi$ angular distribution of particles
relative to the reaction plane angle $\Psi_{R}$ for each nucleus-nucleus collision
and then performs a Fourier decomposition.  
Experimentalists have developed sophisticated techniques for determining the reaction
plane angle $\Psi_{R}$ in each collision~\cite{PHENIX_Whitepaper,PHOBOS_Whitepaper,STAR_Whitepaper}.
The second component $v_2$ of the Fourier decomposition 
is found to be the largest and is referred to as elliptic flow.
\begin{equation}
{{dN}\over{p_{T}dp_{T}dyd\phi}}(p_{T},y,\phi;b) 
= {{dN}\over{p_{T}dp_{T}dy}} 
  \left[1 + 2v_{2}(p_{T},y;b)\cos(2\phi)+... \right]
\end{equation}
One can also determine these coefficients via two-particle correlations 
in $\phi$ and higher order particle correlations, referred to as the 
cumulant technique~\cite{cumulant_method}, which aids in eliminating 
``non-flow'' contributions.

Experiments have extensively measured the $v_2$ coefficient 
for a broad range of hadron species, over a significant kinematic 
range~\cite{PHENIX_Whitepaper,PHOBOS_Whitepaper,STAR_Whitepaper}.  
We emphasize that the experiments do not directly measure pressure gradients 
or flow, but rather the coefficients of the $\phi$ angle distributions.
For example in proton-proton reactions one observes a significant $v_2$ 
coefficient due to hard parton-parton scattering creating opposing jets 
of hadrons.  In heavy ion reactions the this Fourier 
decomposition really measures particle emission directly correlated with 
the orientation of density gradients as demonstrated by
the fact that the $v_2$ for all charged particles 
at low $p_{T}$ scales linearly with the eccentricity of the nuclear overlap region 
(i.e. the exact shape of the ellipse).  
Another striking confirmation comes from two-particle correlations resulting
from Bose-Einstein quantum statistics.  This effect is often referred to as
Hanbury-Brown and Twiss (HBT) correlations~\cite{hbt} and a review of the methodology
as applied to heavy ion reactions can be found in~\cite{hbt_review}.
The STAR experiment has measured the source emission size 
via HBT as a function of the orientation of the two particles relative to 
the reaction plane~\cite{star_hbt_rp}.  These results confirm the spatial deformation 
consistent with expansion from the original ellipse shape.  

\begin{figure}%
\centerline{\psfig{figure=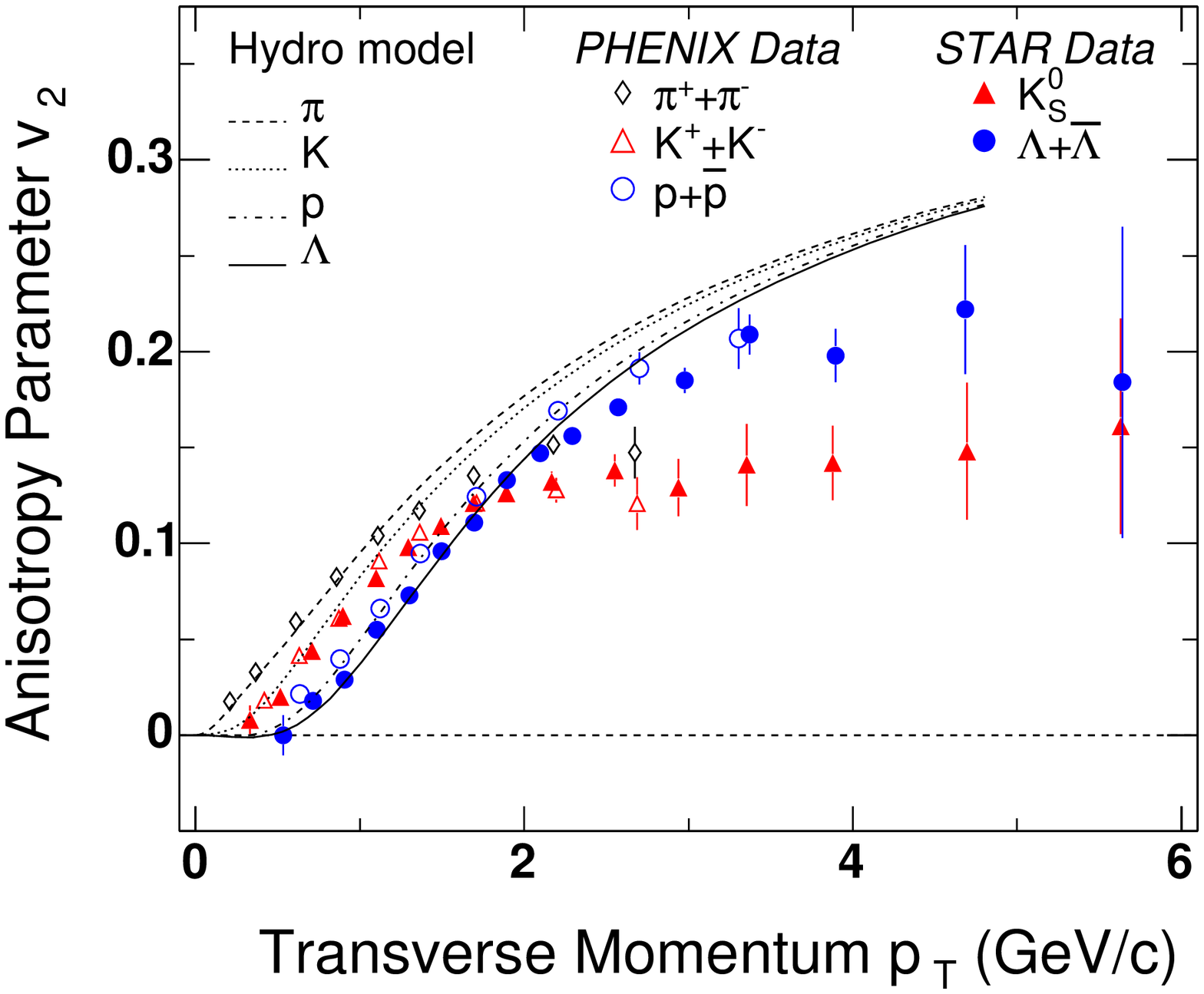,height=25pc}}
\caption{Azimuthal anisotropy ($v_2$) as a function of $p_{T}$ from minimum
bias gold-gold collisions~\cite{STAR_v2,PHENIX_Whitepaper}.  Hydrodynamic calculations are shown
as dashed lines.}
\label{fig_v2_all}
\end{figure}

\begin{figure}%
\centerline{\psfig{figure=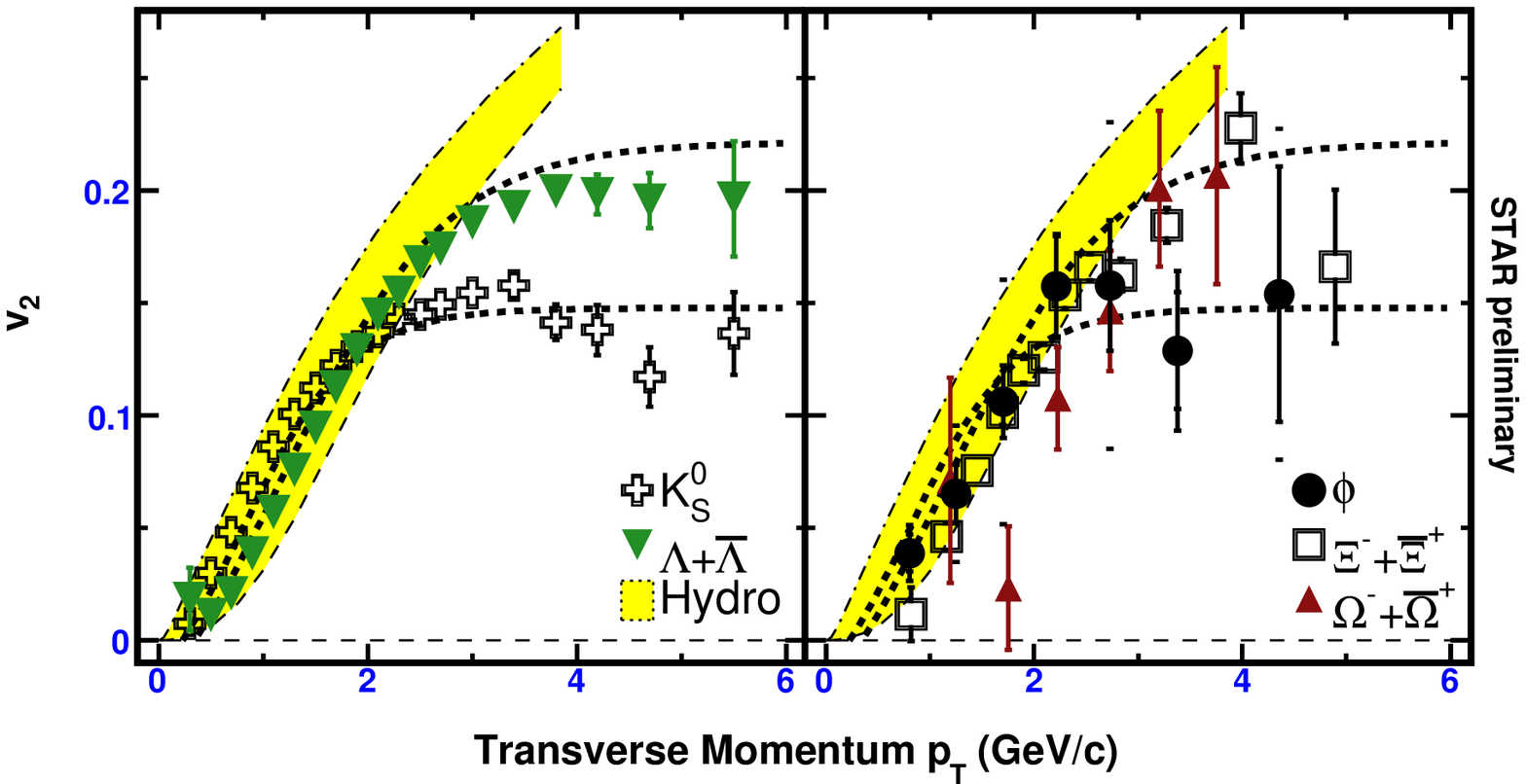,height=20pc}}
\caption{Azimuthal anisotropy ($v_{2}$) for strange (left) and multi-strange (right)
hadrons from the STAR experiment~\cite{STAR_v2QM}.  The curves are empirical fits.  The shaded areas are ranges of
hydrodynamic calculation results.}
\label{fig_v2_multistrange}
\end{figure}

In Figures ~\ref{fig_v2_all} and ~\ref{fig_v2_multistrange} we show $v_2$ as a function of $p_T$ for different 
identified hadron species.  At low $p_{T}<2$~GeV there is a striking 
splitting in the $v_{2}$ pattern where heavier particles have a smaller 
$v_{2}$ than lighter ones at a fixed $p_{T}$. This splitting has a 
natural explanation if the pressure causes the particles to be boosted 
with a common velocity, as is the case in a fluid dynamical picture.  
Also, the increase of $v_{2}$ with $p_{T}$ for all particles is expected 
as higher momentum particles have been given a larger velocity boost and 
are thus more strongly correlated with the maximum pressure direction.
The exact pattern of splitting is sensitive to the equation of state of 
the fluid in this picture and may thus allow for an extraction of the 
equation of state to compare with lattice QCD results.  
An important additional measurement from the 
PHOBOS experiment indicates that $v_{2}$ drops significantly for charged 
particles as one moves away from midrapidity~\cite{phobos_flow}.  This observation contradicts 
many hydrodynamic calculations invoking the Bjorken scenario~\cite{bjorken_scenario}. 
Additionally some have interpretted this as evidence for incomplete 
equilibration and thus a breakdown in the hydrodynamic assumption~\cite{blaizot_inc}.


Measurements of two-particle (HBT) correlations provide significant additional constraints on
the final space-time configuration calculated in fluid dynamical models.  The experimental
data~\cite{PHENIX_Whitepaper,STAR_Whitepaper} reveal significant disagreements with these models when
they match the $v_{2}$ and $p_{T}$ spectra.  This is often referred to as the ``HBT puzzle.''  
Potential solutions to this puzzle are discussed in Section~\ref{sect3a}.

The spectra of even the heaviest hadrons, including the $\phi, \Xi, 
\Omega$, exhibit the same flow-like patterns as the lighter hadrons~\cite{star_multistrange_flow}.  
In the case of the $\phi$ meson, its hadronic cross section with $\pi,N$ 
is small due to OZI rules, which state that processes 
involving diagrams with disconnected quark lines are suppressed~\cite{OZI}. 
One estimate of the inelastic cross section  with nucleons yields $\sigma_{\phi N} \approx$ 9 mb~\cite{egusa}.
The $\Omega$ also has a small 
inelastic cross section, in part due to the lack of $\pi$-$\Omega$ resonant excited states~\cite{vanhecke_omega}, 
but its value $v_{2}^{(\Omega)}$ is comparable to $v_{2}^{(N)}$.  
These small cross sections result in mean free paths for these hadrons that are
large compared to the system size.
Thus, in a hadronic scenario one might have expected the $v_{2}^{(\phi)}$ to be lower than 
$v_{2}^{(N)}$ despite their similar mass.  
Models that attempt to describe the physics in terms of scattering processes 
among quasi-free hadrons (such as RQMD, UrQMD, HSD) predict $v_{2}$ 
values for all hadrons at modest $p_T$ about five times lower than those 
measured at RHIC~\cite{hadron_models_v2}. 
Some have argued that elastic scattering and $3 \rightarrow 2$ inelastic processes~\cite{carsten} 
not included in all hadronic models may significantly increase the expected flow.  
However, the hadronic densities required such that $3 \rightarrow 2$ processes dominate indicates
that multiple hadrons occupy the identical physical space-time coordinates.

This indicates that the constituents of 
the matter during the early time period ($\tau<2$~fm/c), during which
the elliptic flow pattern is established, are not ground state hadrons 
interacting with their standard hadronic cross sections. 
A correlated consequence is that
if the $\phi$ and $\Omega$ whose hadronic production cross sections are OZI 
suppressed are actually formed by the
``coalescence'' or recombination of $s\overline{s}$ and $sss$ respectively, 
then we should expect their production to exactly match
the chemical equilibration model, which the data confirm.

As shown in Figure~\ref{fig_v2_multistrange} for $p_{T}>2$~GeV, the pattern of $v_{2}$ 
for various hadrons appears to 
break down, and baryons have a $v_{2}$ larger than mesons -- in contrast to  
the behavior at low $p_{T}$.  Given the finite size and lifetime 
of the fireball, only a limited number of interactions between constituents 
can occur and thus a finite number of $p_T$ kicks.  
This implies that above some $p_{T}$ the continuous fluid 
approximation must break down.  Naively, one might expect the $v_{2}$ 
values to saturate, as the data indicate, but the picture is complicated
by the observation that the baryons and mesons saturate at significantly 
different values of $v_{2}$.  Again, this may be a confirmation that these 
hadrons are formed from recombination of partons which themselves are the 
original carriers of the flow anisotropy.  

A simple model can be formulated where the flow is carried by quark-like 
constituents (denoted as ``$Q$'') with baryon number 1/3.  In this case, 
if the formation of hadrons results afterwards by coalesence of these 
``quarks'', baryons on average have three times the $p_{T}$ of the 
constituents and $v_{2}(B) = 3 v_{2}(Q)$. The same reasoning applied to 
mesons, with the presumption that a meson is formed by coalescence of a
``$Q\bar{Q}$'' pair, predicts that mesons have on average twice the
momentum of the constituents and $v_{2}(M) = 2 v_{2}(Q)$.  It is tempting 
to identify these medium constituents with the ``constituent'' quarks of
the naive quark model.  However, this is potentially misleading since we 
have no experimental information on their mass or composition.  Nonetheless, 
for the model to work, the medium constituents must have the same quantum 
numbers as the valence quarks of the emitted mesons and baryons.

\begin{figure}%
\centerline{\psfig{figure=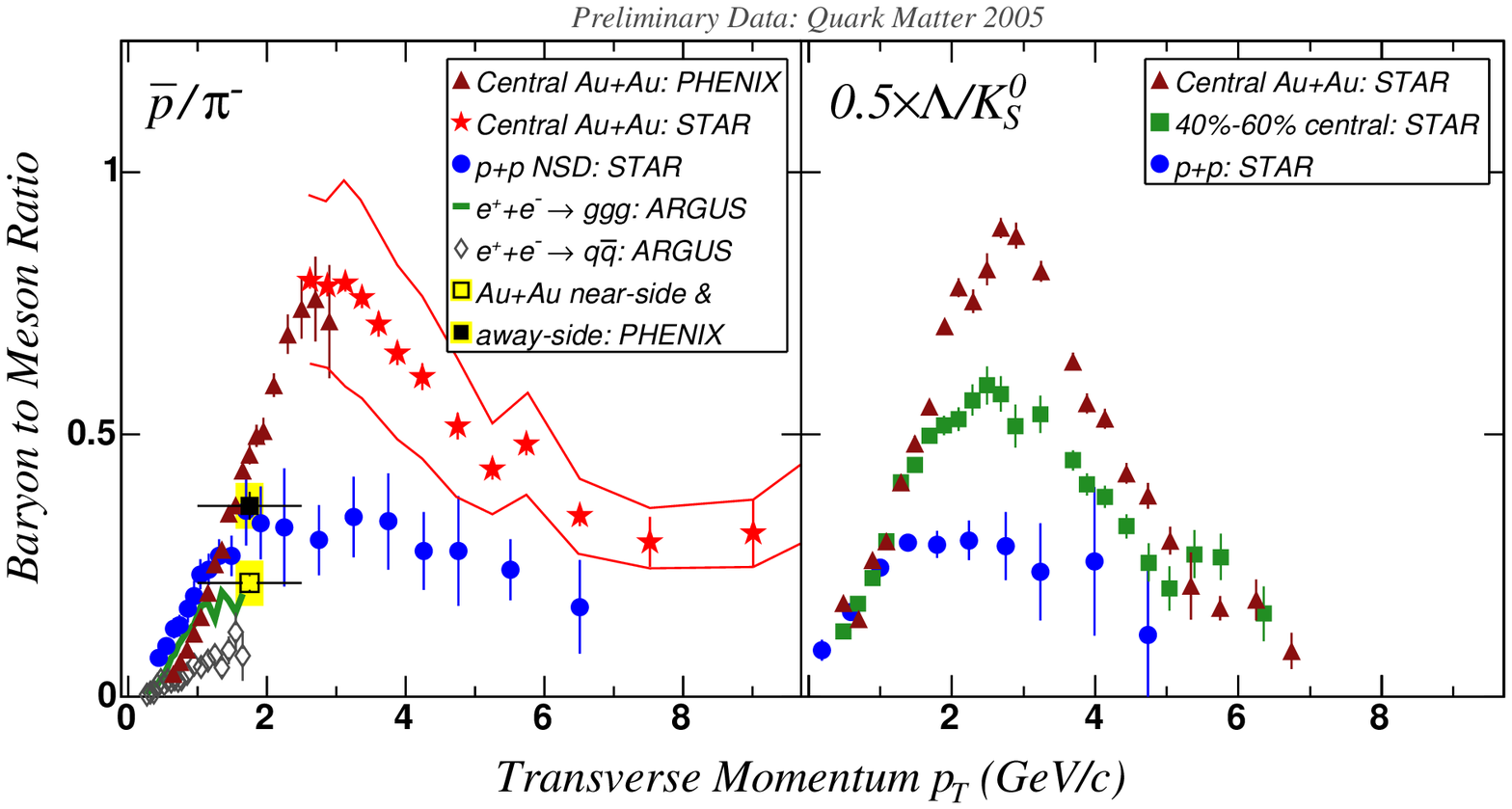,height=20pc}}
\caption{
(Left) The $\overline{p}/\pi^{-}$ ratio at mid-rapidity for central gold-gold
and proton-proton collisions.  (Right)  The $\Lambda/K^{0}_{S}$ ratio in central
and mid-central gold-gold collisions and minimum bias proton-proton collisions.  Values as scaled by 0.5.}
\label{fig_baryon_meson}
\end{figure} 

Another experimental piece of information adding to the so-called ``baryon puzzle'' 
comes from the relative yields of baryons and mesons in the intermediate 
momentum range 2 $< p_{T} <$ 6 GeV/c.  The ratios of hadron yields,
which are dominated by $p_{T}<1$~GeV/c, are well described by the chemical 
equilibrium model.  At the highest $p_{T}>5-8$~GeV/c, the hadron yields 
appear to be consistent with jet fragmentation functions as determined 
from other hadronization processes.  For example, in quark and gluon jets 
measured in proton-proton collisions the $\overline{p}/\pi^{-}$ ratio is about 0.2~\cite{pp_ratios}.  
However, at RHIC in central gold-gold reactions, the 
$\overline{p}/\pi^{-}$~\cite{phenix_baryon_ratio} and similarly 
$\overline{\Lambda}/K$~\cite{star_baryon_ratio} ratios approach unity in the 
intermediate $p_{T}$ range, as shown in Figure~\ref{fig_baryon_meson}~\cite{sorensen}.  
At the highest transverse momenta, even if 
partons propagating through the medium have large in-medium interactions 
(as detailed in Section~\ref{sect2c}), the leading parton is expected to 
punch out of the system before fragmenting in the vacuum.  
Thus, we may expect the ratio of 
hadrons above some $p_{T}$ threshold to be universal.  At more modest 
$p_{T}$, mass related formation time differences between baryons and 
mesons might result in significant deviations from the universal vacuum 
fragmentation ratios.  However, the data do not support a picture which 
associates the peculiarities of the hadron ratios at intermediate $p_{T}$ 
at RHIC with hadronic mass differences.  For example, the $\phi$-meson 
appears to follow a scaling from mid-central to central gold-gold collisions 
similar to the lighter mesons, rather than the equally massive nucleons~\cite{phi_scaling}.  
The simple model of recombination of quark-like 
constituents may also reconcile this part of the puzzle.  Three of these 
constituents need to coalesce to form baryons and only two to form mesons.  
This mechanism effectively boosts baryons to larger $p_{T}$ than mesons, 
and since the $p_{T}$ spectra are rapidly falling, the yields of baryons 
are enhanced compared with meson yields at the same $p_{T}$.  

These observations led the authors of \cite{Fries:2003vb} to reason that 
the ``(recombination) scenario requires the assumption of a thermalized 
partonic phase characterized by an exponential momentum spectrum. 
Such a phase may be appropriately called a quark-gluon plasma.''
However, other observations do not fit into such a simple picture.
If the emission of the intermediate $p_{T}$ hadrons would, indeed, 
occur by recombination from a well thermalized medium, these hadrons 
should not exhibit jet-like angular correlations.   
Experiments have made first measurements of such di-hadron correlations 
and found significant jet-like correlation for both, mesons and baryons~\cite{baryon_jetcorr}.  
This finding lends support to speculations that hadrons in this $p_{T}$ 
range are formed by a combination of ``shower partons'' from fragmenting 
jets and ``thermal partons'' from the medium.  These models need to 
be further tested with very high statistics data for identified particles.  

In the first few years of RHIC running, it was assumed that $c$- and 
$b$-quarks were too massive to participate in collective flow.  
(Since most of their mass is of non-QCD origin, it is retained when 
chiral symmetry is approximately restored in the medium.)  In a perfect
hydrodynamical medium with zero mean free path, all particles must move 
with the same velocity field.  However, the medium created in heavy ion 
reactions lasts only $10-20$~fm/c.  Since the average thermal momentum of
a heavy quark with mass $M\gg T$ is $\langle p^2\rangle \approx 3MT$, it 
would require many interactions with momentum transfer of order $T$ to 
thermalize a $c$- or $b$-quark and let it participate in the collective
flow.  However, recent data from the PHENIX experiment suggest that,
contrary to the expectations, this may in fact occur~\cite{nagle_charmflow}
indicating even stronger interactions in medium.  

One method of extracting heavy meson contributions is from the measurement of ``non-photonic'' 
electrons that are expected to originate mainly from semi-leptonic decays 
of $D$ and $B$ mesons (for example $D \rightarrow K + e + \nu$).  Electrons 
at low $p_{T}$ carry little kinematic information about the parent $D$ meson, 
but for electrons with $p_{T}>1$~GeV/c the azimuthal direction of the electron 
is well correlated with the angle of the parent $D$ meson and the electron 
carries on average 60\% of the $p_{T}$ of the meson.  The measurement is 
challenging since one must first measure the inclusive electron anisotropy 
and then subtract off the anisotropy resulting from ``photonic'' background 
(for example $\pi^{0} \rightarrow \gamma e^{+}e^{-}$).  

The PHENIX experiment has measured the $v_{2}$ coefficient for these 
``non-photonic'' electrons and deduced a significant anisotropy for $D$ 
mesons~\cite{phenix_charm_flow}.  More precision data is needed to determine
if these heavy quarks are flowing with the identical velocity field as the
light partons.

%% file: section2c.tex
As discussed in Section~\ref{sect1b}, the RHIC experiments have opened 
the door on using hard probes to study the medium created in heavy ion 
reactions.  Before using these as probes, one must experimentally understand 
the initial hard scattering processes involved.  

Measurements addressing the question whether
modifications to hard scattering processes observed in nucleus-nucleus 
collisions are significantly influenced by initial-state effects were 
made using deuteron-gold reactions.  In these reactions one 
could still have a depletion of jets or high $p_{T}$ particles due to 
gluon saturation in the incoming gold nucleus, but since no substantial medium 
is created, there should be little chance for the outgoing partons to be 
modified as might occur in a nucleus-nucleus reaction.  Results from all 
four experiments~\cite{dau_all} conclusively show that hadrons at midrapidity with 
$p_{T}\approx 2-10$~GeV/c are not suppressed but rather show a modest 
enhancement, possibly due to initial and final state partonic multiple 
scattering (often referred to as the ``Cronin effect''~\cite{cronin}).  
An important additional observation
was made that at more forward rapidities, sensitive to even lower $x$ partons 
in the gold nucleus, experiments observe a suppression of hadron 
production at modest $p_{T}\approx 2-5$~GeV/c~\cite{dau_forward}.  This effect is currently 
under intense investigation to see if it relates to gluon saturation or 
the CGC. 

A second relevant observation is of direct photon 
production in gold-gold reactions.  The direct photons are all photons emitted 
in the reaction excluding those from hadronic decays of $\pi^{0}, \eta$, etc.  
At low $p_{T}<3$~GeV/c there may be significant direct photon contributions 
from thermal radiation by the medium (see Section~\ref{sect4b}).  
At higher $p_{T}$ we expect the direct 
photons to be dominantly from initial quark-gluon hard scattering processes.  
The PHENIX experiment has measured the yield of direct photons at midrapidity
for $p_{T}= 2-14$~GeV/c in proton-proton reactions and finds the 
results in excellent agreement with NLO pQCD predictions~\cite{phenix_photons}.  Because the 
final-state interactions of energetic photons are small, the yield of  
direct photons in gold-gold reactions can be described by multiplying the yield
measured in proton-proton reactions with the ratio of the incident parton
flux of two gold nuclei to that of two protons.  This normalization is usually 
referred to as {\em binary collision scaling}. The data show the measured 
direct photon yield in gold-gold to be in perfect agreement with this expectation~\cite{phenix_photons}.
Thus, even in these violent gold-gold reactions, factorization and universality 
seem to hold, with the emitted photon passing through the created medium 
unaffected.  The fact that the parton flux of the incident gold nuclei is 
obtained by geometric superposition of the nucleon parton distribution 
functions (PDF) also demonstrates 
that nuclear modifications of the PDF are modest in this kinematic domain.
We note that in the case of this measurement, we are not checking the 
fragmentation function (FF) factorization assumption since the photon is not strongly interacting.

\begin{figure}
\centerline{\psfig{figure=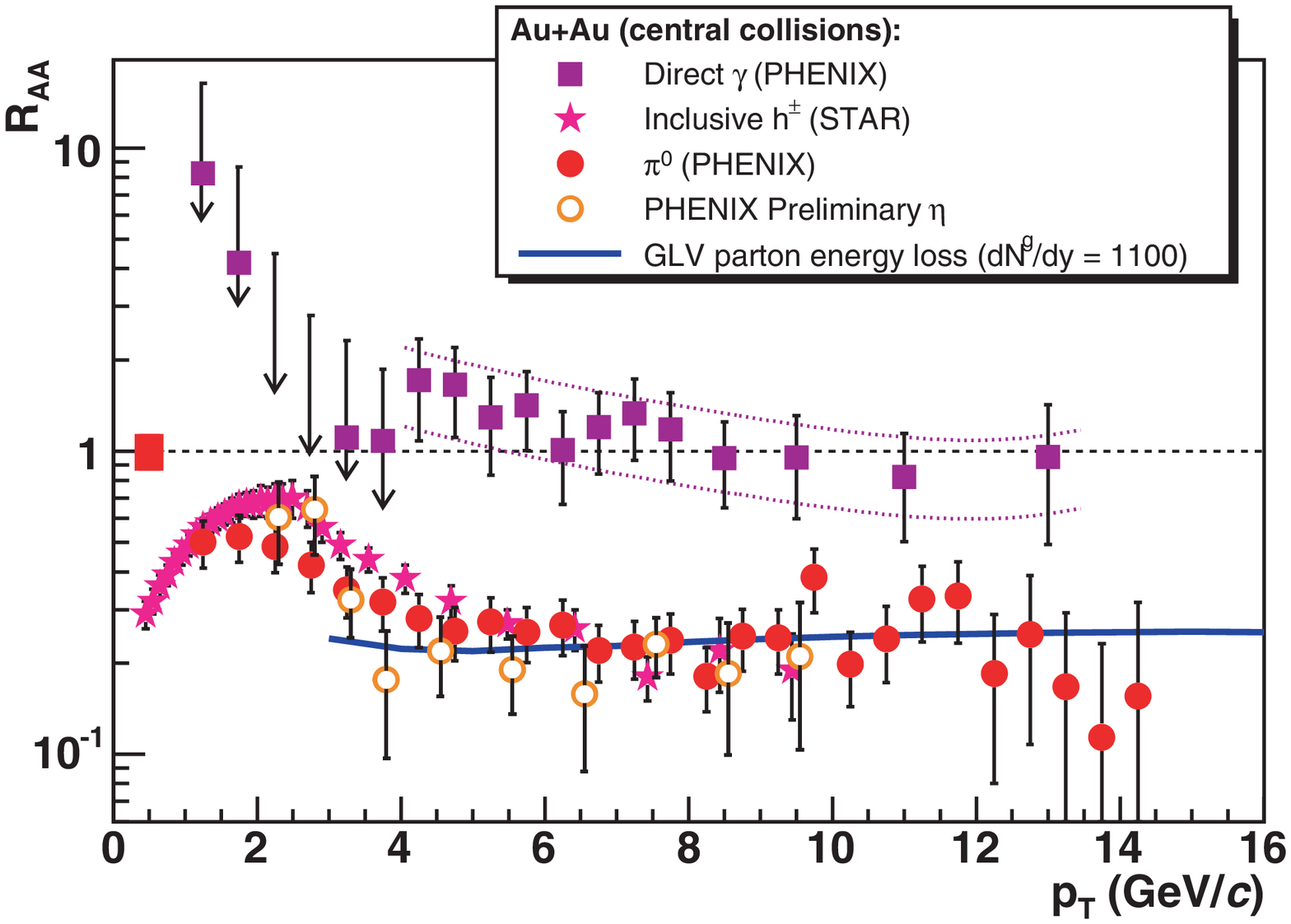,height=25pc}}
\caption{Nuclear modification factor $R_{AA}$ for direct photons and hadrons
in central (0-10\% of $\sigma^{Au+Au}_{inel.}$) gold-gold collisions
at $\sqrt{s_{NN}}=200~$GeV~\cite{reygers}.}
\label{fig_raa}
\end{figure}

After these tests of the calibration framework for hard scattering processes, we turn to 
the observations relevant to quark and gluon probes of the medium.  
In the first data taking at RHIC, the experiments observed 
jet quenching in the form of a suppression in the yield of high-$p_{T}$  
hadron production at midrapidity~\cite{BRAHMS_Whitepaper,PHENIX_Whitepaper,PHOBOS_Whitepaper,STAR_Whitepaper}.  
Shown in Figure~\ref{fig_raa} is the nuclear 
suppression factor $R_{AA}$, which is the measured yield of hadrons 
relative to the expected yield from proton-proton reactions scaled 
according to the binary collision assumption, as a function of $p_{T}$ for 
unidentified and identified hadrons.  
In contrast to the direct photon data, we 
observe a suppression by a factor of five for these hadrons.
The observed suppression can be 
viewed as a modification of the fragmentation functions of quarks and gluons due to 
the surrounding medium.  The detailed interpretation of
these results is discussed in Section~\ref{sect3b}.

One method for isolating quark jets (as opposed to gluon jets) is to look for $D$ and $B$ mesons 
that predominantly come from the fragmentation of charm and beauty quarks. 
Because of their large mass $M$ these heavy quarks are produced
early (on a time scale $1/2M$) and later production by thermal processes 
is strongly suppressed.  There are two methods of measuring heavy flavor 
mesons at RHIC.  Direct D meson reconstruction is presently limited to
low $p_{T}$, but gives useful information on the scaling of the total 
charm cross section~\cite{star_charm}.  These results are consistent with binary 
collision scaling of total charm production, but still have large errors.
Since charm, once created, in unlikely to be annihilated, we expect the 
total number of $c$ and $\overline{c}$ quarks to remain constant after 
their initial production.  The second method involving the measurement 
of ``non-photonic'' electrons has proven to be extremely useful at RHIC.  
Results from the PHENIX and STAR experiment give yields for these electrons 
in proton-proton through central gold-gold collisions over a wide $p_{T}$ range.  
They indicate a significant suppression $R_{AA}\ll 1$ in central gold-gold 
reactions~\cite{elec_raa}.  

This suppression came as a surprise, because expectations were for a 
reduced energy loss for heavy quarks than for light quarks fragmenting
into pions and other light hadrons.  
In the theoretical framework of parton energy loss via gluon radiation,  
this expectation is for two reasons.  First, the D meson almost always 
derives from a charm quark, but the pion spectrum has contributions from 
quark and gluon fragmentation, and the energy loss of gluons is expected 
to be about twice that of quarks.  Second, the more massive charm quark 
should suffer a ``dead-cone'' effect suppressing the forward emission of 
gluons~\cite{Dokshitzer:2001zm}.  The strong suppression of heavy meson production has led to 
a re-assessment of alternative scenarios of parton energy loss, and is
discussed in Section~\ref{sect3b}.

Measurements of jet quenching phenomena at RHIC are not restricted to 
single particle yields.  Since fragmenting partons are the result of 
hard parton-parton scattering processes, they should be accompanied by 
a partner jet at the opposite azimuthal angle.  The RHIC experiments 
are not capable of complete jet reconstruction in heavy ion reactions
due to the modest $p_T < 25~$GeV of the jets and the large background of uncorrelated
low $p_T$ particles in central gold-gold collisions.  Instead, experiments correlate 
identified hadrons in momentum space. The STAR and PHENIX experiments 
have shown analyses that use the highest $p_T$ hadron as the trigger 
particle, defining azimuthal angle $\phi=0$, and then plot 
$\Delta \phi$ with other hadrons (associated particles) in the same event.  
In proton-proton reactions this analysis results in a sharp peak centered 
at $\Delta\phi=0$, due to other hadrons fragmenting from the same parton 
as the trigger particle, and a wider peak near $\Delta\phi=\pi$, due to
hadrons fragmenting from the opposing parton. 

\begin{figure}
\centerline{\psfig{figure=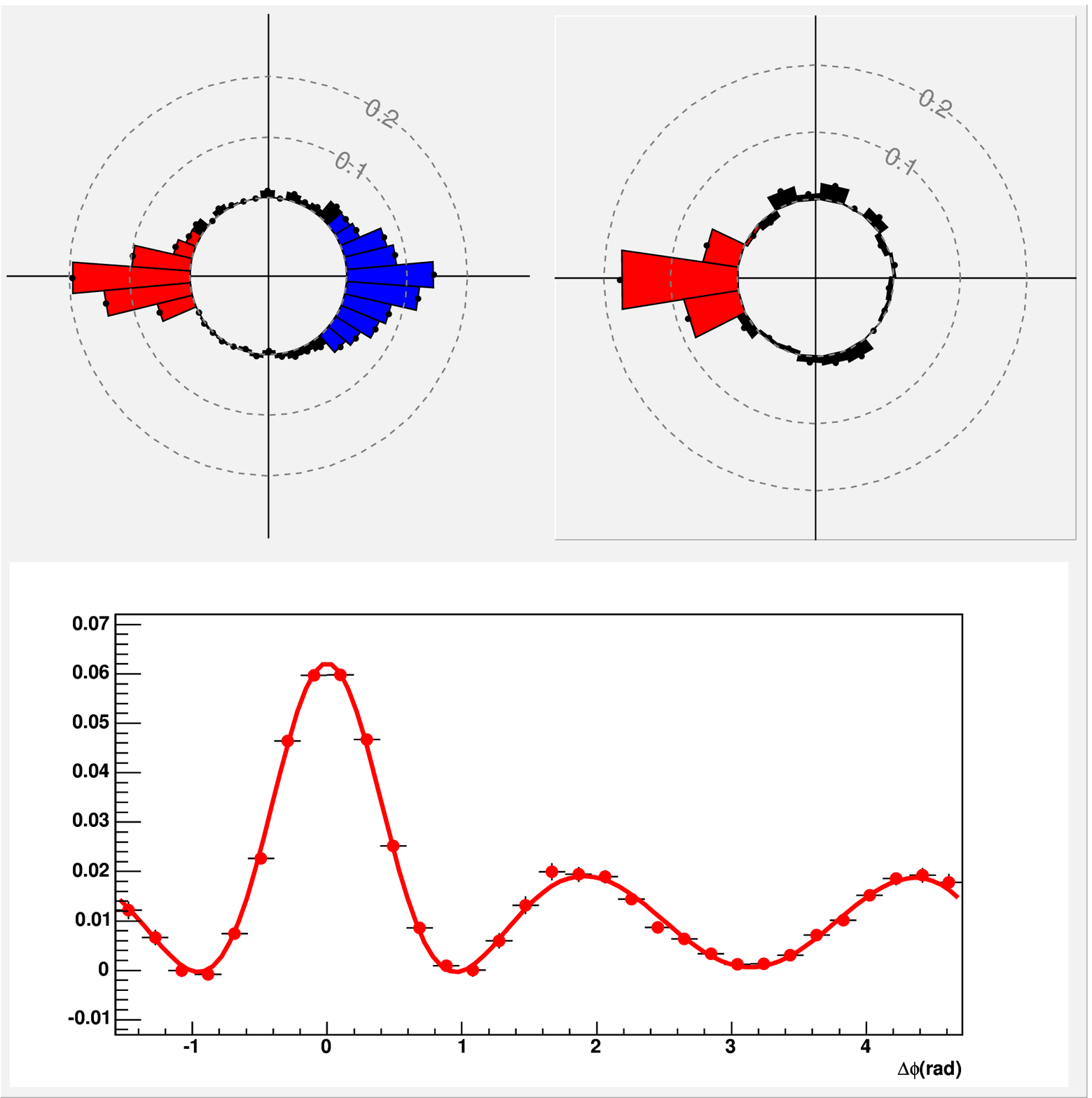,height=25pc}}
\caption{Upper panels results from the STAR experiment are the azimuthal distributions of unidentified hadron pairs
for (left) proton-proton reactions and (right) background subtracted central
gold-gold reactions.  All correlation functions require a trigger particle with $4 < p_{T} < 6~$GeV/c
and associated particles with $p_{T} > 2~$GeV/c.
The lower panel are preliminary results from the PHENIX experiment on the
azimuthal distribution in central gold-gold reactions requiring the trigger particle
$2.5 < p_{T} < 4.0~$GeV/c and the associated particles with $2.0 < p_{T} < 3.0~$GeV/c.}
\label{fig_jetcorr}
\end{figure}

However, in gold-gold reactions, first results from the STAR experiment 
triggering on a high $p_T$ particle and correlating it with associated 
hadrons with $p_{T} > 2$~GeV/c showed the complete disappearance of the 
opposite side peak~\cite{star_correlA}, as shown in Figure~\ref{fig_jetcorr}.  
Energy and momentum must be conserved
globally, implying that particles from the opposing parton must 
either be shifted to lower energy ($p_{T} < 2$~GeV/c) and/or be scattered
into a diffuse angular distribution.  New results from higher statistics
gold-gold data reveal that both effects are significant~\cite{starphenix_correlB}.  In lowering 
the required $p_{T}$ for associated particles to 200 MeV/c, one recovers 
the energy of the opposing parton.  Remarkably these particles are spread 
out over nearly the entire opposite side hemisphere ($|\Delta\phi|>\pi/2$) 
and with a mean $\langle p_{T}\rangle$ not much larger than the average 
$p_T$ of all other particles emitted from the medium~\cite{STAR_Whitepaper}.  The data may
indicate that the peak of the backward angular distribution is no longer
at $\Delta\phi=\pi$, but near $\Delta\phi\approx 2\pi/3$~\cite{henner}, shown in 
the lower panel of Figure~\ref{fig_jetcorr}.  
The experiments have embarked on measuring three- 
and more-particle correlations in order to understand the emission of 
these associated particles in detail.

A different probe of the medium comes from quark-antiquark pairs rather 
than just single partons.  The possible suppression of heavy quarkonia 
($J/\psi, \chi_c, \psi', \Upsilon$) due to color screening -- analogous 
to Debye screening in an electromagnetic plasma -- has long been proposed 
as a probe of deconfinement~\cite{matsui_satz}.  Like in the case of jets, the usefulness 
of heavy quark bound states as probes of dense matter must be calibrated 
against their behavior in proton-nucleus or deuteron-nucleus 
reactions.  Measurements of the absorption of $J/\psi$ in deuteron-gold reactions 
by the PHENIX experiment indicate that the state has a break-up cross section 
with nucleons in the gold nucleus of order $1-3$ mb~\cite{jpsi_dau}.  Note that at this energy
the gold nucleus sweeps over the quark-antiquark pair on a time scale 
of $2R/\gamma c \approx 0.14$~fm/c, i.~e.~before the bound state has been
formed.  

Results from the PHENIX experiment in copper-copper and gold-gold 
reactions reveal a suppression of $J/\psi$ yields relative to binary 
scaling expectations~\cite{hugo}.  The suppression appears to be greater than can
be accounted for by break-up from the two gold nuclei passing through, 
thus leaving some room for additional suppression in the medium created.  
However, the level of suppression is not very different from that observed at lower energies by the NA50 
experiment at the CERN-SPS~\cite{na50}.  Model calculations assuming color 
screening of the $J/\psi$ state in medium predicted a much larger suppression 
at RHIC energies due to the larger parton density, higher temperature, and 
longer lifetime of the system~\cite{Digal:2001bh}.  One possible explanation is that the 
color force responsible for binding the
$J/\psi$ is in fact not strongly screened in the medium, as recent lattice 
results suggest~\cite{Asakawa:2003re}.  Another possibility is that one forms additional 
$J/\psi$ later in the reaction from the coalescence of $c\bar{c}$ quark 
pairs or hadronically via $D\overline{D}\to c\bar{c}$ reactions~\cite{charm_reco}.  
Future detailed measurements of open charm and quarkonia are required to discriminate
between these various scenarios.

\begin{figure}
\centerline{\psfig{figure=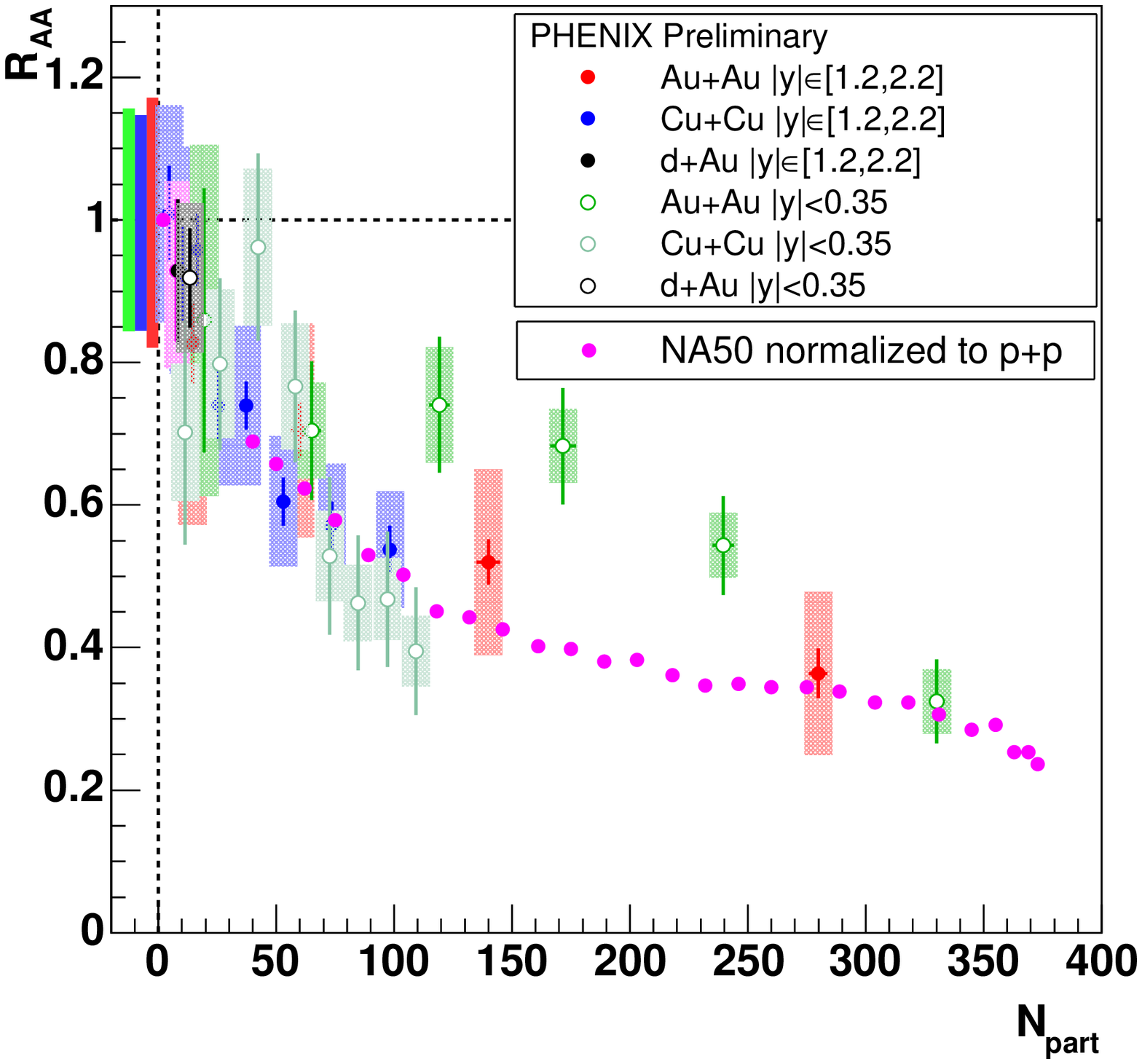,height=25pc}}
\caption{PHENIX preliminary results on $J/\psi$ nuclear modification factor $R_{AA}$ as a function of the number
of nuclear participants~\cite{hugo}.  Results are shown from gold-gold, copper-copper, and deuteron-gold reactions at forward
and mid-rapidity.  Also shown are results from the NA50 experiment from
lower energy at the CERN-SPS~\cite{na50}.} 
\label{fig_jpsi}
\end{figure}

%% file: section3a.tex

As we discussed in Section~\ref{sect2b}, the ``elliptic flow'' parameter 
$v_2$ characterizing the azimuthal anisotropy of the hadronic spectra 
depends on the hadron species and exhibits an almost linear dependence 
on the transverse momentum $p_T$. The strong $p_T$-dependence
suggests that this cannot be a simple consequence of the geometric 
aspect ratio of the emitting source \cite{Huovinen:2001wn}. 
The strong dependence of the magnitude of $v_2$ on the hadron mass 
for small $p_T$, which nicely matches the predictions of fluid 
dynamical simulations of the expansion of an ellipsoidal fireball
\cite{Kolb:2001qz}, suggests that $v_2$ is the result of an 
anisotropic expansion caused by the difference in pressure gradient
in and out of the collision plane. This does not mean that other 
effects, such as an angular dependence of absorption, can be entirely 
neglected, but they are subdominant in the bulk emission region 
($p_T<1$~GeV/c) \cite{Retiere:2003kf}.

The generally good agreement between the $v_2(p_T)$ curves expected
from the collective expansion of an ellipsoidal fireball and the 
measured $v_2$ curves for a variety of hadrons makes it worthwhile 
asking to which aspects of the expansion they are sensitive. The 
most obvious details are: the initial conditions, the equation of 
state, and the viscosity of the expanding matter. The formation of
a collective flow velocity in an inviscid fluid is governed by the 
relativistic Euler equation
\begin{equation}
\frac{d}{dt}(\varepsilon \vec{v}) = - \nabla P ,
\label{eq3a1}
\end{equation}
implying that the formation of collective flow relies on the existence 
of a pressure gradient. Elliptic flow requires the existence of an 
azimuthally anisotropic pressure gradient. Since transverse expansion 
quickly reduces any initially existing anisotropy of this kind, it can
be shown that a substantial elliptic flow in noncentral collisions must 
be formed during the early stages of the fireball evolution
\cite{Kolb:2000sd}. It hence requires the rapid formation ($< 2~$ fm/c) 
of a large pressure by equilibration. 

Quantitative solutions of relativistic fluid dynamics confirm this 
expectation. The elliptic flow is generated mainly during the first 
5 fm/c of the expansion, and the pressure gradient has to become active 
before 1 fm/c \cite{Heinz:2001xi}. 
These results imply that the observed elliptic flow is generated 
when the energy density of the matter exceeds 1 GeV/fm$^3$.
The calculations also show a sensitivity to the equation of state, 
but no consensus has been reached yet as to how well the 
$\varepsilon(P)$ relation can be determined in this way 
\cite{Kolb:2000sd,Teaney:2001av,Huovinen:2005gy}. 

A nonvanishing viscosity modifies the right-hand side of (\ref{eq3a1})
and thus affects the generation of elliptic flow. Numerical solutions
of the relativistic Navier-Stokes equations have shown that even a low
(but non-zero) 
shear viscosity $\eta$ has a dramatic effect on the value of $v_2$ 
generated during the expansion \cite{Teaney:2003pb}. 
The approximations and idealizations necessary presently preclude a 
precise determination of an upper limit for $\eta$ compatible with 
the elliptic flow data, but it seems safe to conclude that the 
condition $\eta/s \ll 1$ must hold.

One problem arising when one tries to pin down the equation of state
by a comparison between the data ($p_{T}$ spectra and $v_2$ and the
space-time information from two-particle-correlations) and 
the results of hydrodynamic calculations is that it is difficult to
obtain a good description of all data simultaneously. In part, this 
problem arises from the failure of fluid dynamics to provide for an
accurate description of the expansion during the final, hadronic gas
phase \cite{Hirano:2005wx}. The hadron gas has a rather large viscosity
and also fails to maintain chemical equilibrium during the cooling 
period until particle freezeout occurs. Hybrid descriptions in terms
of fluid dynamics for the deconfined phase followed by a Boltzmann 
cascade for the hadronic phase will make it possible to address this 
issue \cite{Hirano:2004rs,Nonaka:2005aj}.

An important source of information about the collective flow comes 
from the spectra and angular distributions of heavy mesons. Originally,
heavy quarks were not expected to participate in the collective flow 
to a significant degree, but this expectation has been revised in
view of the observed strong suppression of hadrons containing heavy 
quarks at large $p_T$ (see Section~\ref{sect2c}). A large energy loss 
requires large cross sections for heavy quarks in the medium. One 
would then also expect the heavy quarks to more fully reflect the 
global flow of the medium. Because their production is a hard process
even at moderate momentum, heavy quarks may become more controlled 
probes of the pre-hadronic collective flow than hadrons composed 
solely of light quarks \cite{Lin:2003jy,Greco:2003vf}. The large mass
of heavy quarks may also allow for a more rigorous description of 
the hadronization process and thus make quantitative tests of quark 
recombination models possible \cite{Braaten:2001uu}. We will return 
to the transport properties of heavy quarks in the medium in 
Section~\ref{sect3b}.

%% file: section3b.tex
As explained in Section~\ref{sect1b}, the energy loss of a light parton 
in dense matter is controlled by a transport coefficient $\hat{q}$, which 
is proportional to the density of color charges in the medium.
The value of $\hat{q}$ can be calculated in pQCD, where it is found 
to scale like a power of the energy density $\hat{q}=c\varepsilon^{3/4}$
with $c\approx 2$ independent of the temperature, and even holding for 
cold nuclear matter \cite{Baier:2002tc}. 
However, the analysis of the RHIC jet quenching data, assuming a thermal 
quark-gluon plasma and the boost invariant Bjorken scenario, yields a 
much larger value, corresponding to $c_{\rm exp}\approx 10$ 
\cite{Eskola:2004cr}. If this analysis is correct, it implies a 
five times shorter mean free path for hard partons than is predicted
by perturbation theory.  One can view this result as a confirmation of 
the conclusion described above, that the viscosity of the medium must
be very low, because that also implies a very short mean free path 
for thermal particles.

A slightly different analysis of the jet quenching data assumes that
the energy loss can be described perturbatively and uses the data to
deduce the density of scattering centers in the medium. If these are
all assumed to be gluons, the data require a density $dN_g/dy$ of 
$1000-1200$ per unit rapidity \cite{Vitev:2002pf,Vitev:2004bh}. 
It is notable that these calculations do not 
imply that the medium constituents are objects with open color (e.g. gluons).  In deep 
inelastic scattering reactions the photon can scatter from individual quarks 
if its virtuality is such that its wavelength is shorter than the size of the 
proton (about 1 fm).  Similarly, the hard scattered parton could resolve 
color charges within the constituents of the medium, even if these objects
were color singlets, such as hadrons.  

As discussed earlier, radiative energy loss is less effective for heavy
quarks than for light quarks and gluons. 
As a result, calculations \cite{Djordjevic:2005db} show 
that the suppression of mesons with charm should be only about half as 
strong as the suppression of light mesons in the range $p_{T}<$~10 GeV/c.  
The suppression of $B$-mesons in this range is even 
smaller, $R_{AA}^{(B)}>0.8$, if realistic gluon densities in the
medium are assumed.  These results are in strong disagreement with
the data on ``non-photonic'' electrons.

One possible resolution of this quandary is the realization that 
collisional energy loss is important for heavy quarks \cite{Mustafa:2004dr},
triggering a re-evaluation of its importance 
for light quarks and gluons as well. Since the scattering centers 
in the medium are not infinitely massive, a hard parton loses energy 
even in an elastic collision.  Because the energy loss per collision is
small, but the parton suffers many collisions in the medium if its 
mean free path is small, the collisional energy loss can be described
by a Langevin equation in which the diffusion coefficient $D$ is an 
adjustable parameter \cite{Moore:2004tg}.  
In order to achieve reasonable descriptions 
simultanously for the ``non-photonic'' electron $R_{AA}$ and $v_{2}$,
one needs a diffusion constant in the range $3 < 2\pi TD < 6$, which
corresponds to a very small shear viscosity $1 < 4\pi (\eta/s) < 2$.
Such a small diffusion constant near the conjectured absolute lower 
bound \cite{Kovtun:2004de} is difficult to reconcile with perturbation 
theory.  One possible explanation of a large charm energy loss would
be the presence of $D$-meson like resonances in the medium above $T_c$ 
\cite{vanHees:2004gq}.

Recently a study including both, radiative and collisional energy loss, 
has been made which includes effects of the collision geometry as well
as path length fluctuations \cite{Wicks:2005gt}.  The study uses 
perturbative cross sections and assumes a density of gluons in the medium. 
The authors conclude that the two effects combine to explain the observed 
$R_{AA}$ for light and heavy mesons simultaneously for a realistic gluon 
density $dN_g/dy \approx 1000$.

One problem with these estimates is that they assume that the quark-gluon
plasma is predominantly composed of gluons. If the medium is chemically 
equilibrated, it contains a substantial fraction of quarks and antiquarks,
which have smaller perturbative cross sections than gluons. The required 
rapidity density then rises to about $dN_{qg}/dy = 2000$.  The problem with 
this result is that it is incompatible with the final entropy density 
$dS/dy \approx 5000$ deduced from the measured charged hadron multiplicity 
\cite{Pal:2003rz}. The entropy density at early times cannot be less, 
putting a limit $dN_{qg}/dy \leq \frac{1}{4} dS/dy \approx 1300$.  
This argument again suggests that the parton-parton cross sections in the 
medium must be significantly larger than predicted by perturbation theory.

%% file: section3c.tex
The RHIC data constrain the thermodynamic properties of the matter 
produced in Au+Au collisions, such as its entropy density.
The total final-state entropy can be deduced either from the 
measured hadron yields in combination with information on the 
size of the emission volume, or from the implied hadron yields at
chemical freeze-out assuming full chemical equilibration.  The
first of these methods starts from the expression for the entropy in  
terms of the phase-space distribution $f_i(x,p)$ of non-interacting
particles ($i$ denotes the particle species, $d_i$ its degeneracy):
\begin{equation}
S=\sum_i d_i \int \frac{d^3x d^3p}{(2\pi)^3} [ - f_i \ln f_i
  \pm (1\pm f_i) \ln(1\pm f_i) ] ,
\label{eq3c1}
\end{equation}
where the $+(-)$ sign holds for bosons (fermions). The momentum 
dependence of the functions $f_i(x,p)$ is determined from the 
measured or fitted $p_{T}$ spectra. The spatial x-dependence is assumed to be 
Gaussian, with radii fitted to the two-particle correlations using 
the Hanbury-Brown--Twiss (HBT) effect. This approach yields a final 
entropy per unit rapidity at $y\approx 0$ of $dS_f/dy \approx 4450$ 
for the 11\% most central Au+Au events at $\sqrt{s_{NN}} = 130$~GeV 
\cite{Pal:2003rz}. This number scales to $dS_f/dy \approx 5600$ at 
midrapidity in the 6\% most central events for the highest RHIC 
energy ($\sqrt{s_{NN}} = 200$~GeV).

The second method starts from the entropy per particle $S/N=7.25$ 
of a fully equilibrated hadronic resonance gas at the chemical 
freeze-out temperature $T_{\rm chem}\approx 170$~MeV \cite{Nonaka:2005vr}. 
Using the fit of hadron yield ratios and the total measured charge 
multiplicity, one can deduce an entropy content of $dS_{\rm chem}/dy 
\approx 5100$ at chemical freeze-out \cite{Muller:2005en}. 
Since the cooling from $T_{\rm chem}$ until final decoupling of all 
hadrons at $T_f\approx 110$~MeV proceeds through the rather viscous 
hadronic gas phase, the implied increase of the entropy by 10\% 
between chemical and kinetic freeze-out is realistic \cite{Hirano:2005wx}. 
On the other hand, as discussed in Section~\ref{sect3c}, the 
collective flow analysis tells us that the viscosity at even earlier
times must be very small.  We can thus conclude from the RHIC data 
that the entropy during the hydrodynamic expansion of the matter at
temperatures $T>T_c$ is $dS/dy\approx 5000$ with an estimated error 
of $\pm 400$.  In the framework of three-dimensional relativistic 
fluid dynamics, this result can be used to obtain estimates of the
evolution of the entropy density of the matter between the moment 
of first equilibration and the transition into a hadronic gas at $T_c$.
Assuming a boost invariant longitudinal expansion and averaging over
the transverse plane, one obtains an entropy density of about 33/fm$^3$
at a time of 1 fm/c \cite{Muller:2005en}, corresponding to a temperature
of 270 MeV for 30 massless degrees of freedom. This estimate assumes 
that the entropy growth during the ``ideal liquid'' phase of the medium 
is negligible.

It thus appears that a rather precise determination of the entropy 
density of the hot medium at early times with an uncertainty in the 
range of $(5-10)$\% is within reach, once an analysis without the 
assumption of boost invariance has been performed. 
If it were possible to obtain a similarly
accurate measurement of the energy density, from jet quenching data 
or other considerations, this would enable a model independent 
determination of the equation of state \cite{Muller:2005en}, which 
could be compared with the lattice predictions. In particular,
it would allow for a determination of the effective number of (massless)
degrees of freedom in the medium. Lattice QCD predicts that this number 
rises steeply near $T_c$, but then remains almost constant over a wide
temperature range. This prediction could then be tested by comparing 
results obtained at widely different beam energies.

%% file: section3d.tex
Perhaps the most direct source of evidence for the nature of the 
matter produced at RHIC is derived from the recombination model 
of hadron production at intermediate $p_{T} = 2-5$~GeV/c
\cite{Greco:2003xt,Fries:2003vb,Greco:2003mm,Fries:2003kq}. In this
model, which gives successful descriptions of measured hadron yields 
and $v_{2}$ in terms of an underlying thermal valence quark spectrum,
the primary constituents of the medium just prior to hadron emission
have the quantum numbers of quarks and antiquarks in complete chemical
equilibration.  The theoretical argument for rapid flavor equilibration 
of $u,d,s$ quarks is based on the abundance and large cross section of 
thermal gluons at temperatures above $T_c$ in perturbative QCD
\cite{Rafelski:1982pu}.  This 
expectation is confirmed by the near perfect flavor equilibration of
the emitted hadrons.  However, this raises the question: what happens to
the thermal gluons as the system cools through $T_c$?  

In part, the answer may be that the gluons get incorporated into
the emitted hadrons. In the recombination model, one can show that
the emission probability of a hadron from a thermal medium does 
not depend on details of its internal wave function
\cite{Muller:2005pv}. If the wave
function contains higher Fock states including additional gluons 
besides the valence quark configuration, the emission process can
absorb some of the gluons in the medium. However, the internal 
structure of the emitted hadrons does have an influence on the 
valence quark scaling law for the $v_{2}$ component of the collective
flow. Components of the hadron wave function containing additional 
gluons or sea quarks lead to deviations from the naive scaling law 
in the few percent range which could be detected by precise measurements. 

A great deal of the uncertainty in our understanding of the process
of hadronization is rooted in our lack of precise knowledge of the
structure of the quark-gluon plasma in the temperature region just
above $T_c$. Various results from the lattice, as well as the RHIC
data, indicate that it is a strongly coupled state.  As we briefly 
mentioned in Section~\ref{sect1b}, it is possible to use lattice 
results for the susceptibilities 
of various quark flavor related quantities, such as strangeness, electric
charge, and baryon number, to conclude that the excitations of the 
matter just above $T_c$ have the quantum numbers of individual quarks,
not diquarks or quark-antiquark pairs \cite{Koch:2005vg,Ejiri:2005wq}.  
The analysis shows that the transition from a medium dominated by meson 
($q\bar{q}$) and baryon ($qqq$, $\bar{q}\bar{q}\bar{q}$) bound states 
to quark-like states occurs rapidly and completely when the temperature 
exceeds $T_c$ \cite{Majumder:2006}.  The best way to test this regime 
may be with hadrons containing heavy quarks, for which the hadronization
process is much better understood (see e.~g.~\cite{Braaten:2001uu}).

%% file: section4a.tex
There have been two previous experimental programs with 
relativistic heavy ion collisions at energies sufficient 
to consider the possible formation of a quark-gluon plasma. 
Both were fixed target programs. One used the Alternating 
Gradient Synchrotron (AGS) at Brookhaven National Laboratory 
to accelerate heavy ion species up to gold nuclei and center-of-mass 
energies per nucleon pair in the range 
$\sqrt{s_{NN}}=2.4-4.8$~GeV and the other the Super Proton 
Synchrotron (SPS) at CERN with species up to lead nuclei with 
$\sqrt{s_{NN}}=7.6-17.3$~GeV.

The BNL-AGS program resulted in strong evidence for the 
creation of ``resonance matter'' where most all hadrons are in 
excited states cascading and decaying into 
stable hadrons~\cite{resonance_matter}.
The possiblity of modest temperature and finite baryon density 
quark-gluon plasma formation in the very short earliest stages 
is not ruled out, but the program lacked measurements of 
penetrating probes of these early time stages. Intriguing 
results such as enhanced $\overline{\Lambda}/\overline{p}$ 
persist~\cite{antilambda_puzzle}, but are currently unexplained.

The CERN-SPS program had a broader suite of experimental 
observables and was at significantly higher collision energy.  
In the year 2000, there was a CERN press release announcing 
``evidence for a new state of matter''  In evaluating the
experimental evidence they concluded that ``the new state of 
matter found in heavy ion collisions at the SPS features many 
of the characteristics of the theoretically predicted quark-gluon 
plasma'' \cite{Heinz:2000bk}.\footnote{
Since there was no peer-reviewed publication associated with this
announcement, we refer here to the document that accompanied the press release.}
It is notable that the document describes the quark-gluon plasma as 
a state where ``quarks and gluons... freely roam within the volume 
of the fireball''.\footnote{The authors have clarified \cite{Heinz:2006pc} 
that ``roaming'' was not intended to mean moving without interactions.}
The argument placed significant weight 
on the ``strong nonlinear dependence'' of the enhancement 
of multi-strange hadrons and the suppression pattern of charmonium 
states where ``intermediate $c\overline{c}$ seems to disappear quite 
suddenly in semicentral lead-lead collisions, and in the most central  
lead-lead collisions an additional reduction of the $J/\psi$'' is 
observed.  

Over the past five years, changes in the experimental data and 
theoretical understanding have tempered some of these conclusions.  
First, it is quite unlikely that at CERN-SPS energies or even at 
RHIC, the quarks and gluons should ``roam freely.'' While quarks 
and gluons are most likely liberated from their hadronic bonds,
they are still strongly interacting in the medium with mean free 
paths less than 1 fm.  Second, the concept of ``sudden'' observed 
transitions as a function of energy density is recognized as 
unnatural even in the case of a first-order phase transition, 
where observables exhibit discontinuities as function of temperature,
but not as function of the energy density. Given the evidence for 
a smooth cross-over transition from lattice calculations, such 
discontinuities appear highly unlikely.  Indeed, the experimental 
evidence for ``non-monotonic derivatives'' in the suppression of 
$\overline{c}c$ states~\cite{na50nonmono} has not held up to further 
scrutiny, and strangeness enhancement appears to have a gradual onset 
with significant effects already seen in proton-nucleus reactions
\cite{cole,Susa:2002rf}.  

Nonetheless, these lower energy fixed target programs provided 
major advances in the field. Many of the experimental observations 
still challenge our theoretical framework and emphasize the 
importance of measuring the excitation function (collision energy 
dependence) of many observables from the AGS to the CERN-SPS to 
RHIC and, in the future, to even higher energies in the LHC heavy 
ion program.  
Not only is the excitation function of interest, there is unique physics at
these lower energies.  The CERN-SPS program on heavy quarks and low
mass dileptons has been revived with the NA60 experiment~\cite{na60}.  The NA49 experiment
has mapped out hadronic observables over the complete energy range between
the BNL-AGS and the top CERN-SPS energy~\cite{na49}, and the NA45 experiment
has focussed on low mass dileptons and also higher $p_T$ hadrons~\cite{na45new}.  
At these lower energies, 
matter at the highest net baryon density will be studied with precision experiments 
at the planned FAIR facility at GSI~\cite{fair_gsi}.  Although at RHIC the transition to 
a QGP appears to be a smooth cross-over, there may exist a tri-critical point in the phase diagram at
large net baryon density where the transition changes to first order.  
These lower energy programs have a unique opportunity
to study this region of the phase diagram of nuclear matter.

%% file: section4c.tex
The great progress in characterizing and understanding the medium created 
at RHIC over the past five years invites the question of what are the critical
outstanding issues and key additional measurements needed to further 
characterize the properties of the quark-gluon plasma.  

The four experiments have made an almost complete determination of the 
global features of the medium and its collective motion at 
$\sqrt{s_{NN}}=200$ GeV.\footnote{It 
should be noted that this level of detailed measurement remains to be 
completed at lower energies all the way down to RHIC injection energy, 
where one can make direct comparisons to the fixed target CERN-SPS results.} 
The main experimental thrust of the future will be in the area of hard 
and rare probes with low cross section or small branching fractions,
which require an increased luminosity and enhanced detection capabilities.

In order to better dissect the mechanism responsible for jet quenching 
and more accurately determine the color charge density in the medium,
we want to distinguish quarks traversing the medium from gluons.  
The color octet gluons should lose energy at roughly twice the rate
as the color triplet quarks.  The present measurements of high $p_{T}$ 
hadron suppression cannot separate contributions from quark and gluon 
fragmentation, and the standard techniques used to isolate gluon jets 
via 3-jet event analysis do not work for modest $p_T$ jets in the high background of 
nucleus-nucleus collisions at the RHIC beam energy.  
In the future, RHIC experiments will use identified hadrons to separate 
these jet contributions on a statistical basis.  One prediction is that 
the $\overline{p}/p$ ratio should drop at high $p_{T}$ since antiprotons
are statistically more likely to originate from a gluon initiated jet, 
which should lose more energy in the medium.  

An alternative method of separating quark jets is by tagging charm and 
beauty mesons at high $p_T$, where they are almost exclusively fragmentation 
products of heavy quarks. Both PHENIX and STAR have proposed upgrades 
including inner silicon detectors to measure displaced vertices from 
these heavy flavor decays.   Not only does this tag quark jets, but 
also allows for the study of the ``dead cone'' effect due to their finite
velocity if radiation is the dominant energy loss mechanism.  Direct reconstruction
of charm mesons and baryon states, tagging single electrons from D and B meson 
decays, and tagging $J/\psi$ from B decays will give experiments many handles
on this physics.

Further details can be gleaned by studying $\gamma-jet$ correlations since 
the photon tags the jet energy at leading order in $\alpha_s$.  This method 
requires higher beam luminosity, which can be achieved through an accelerator
upgrade including electron cooling of the gold ion beams.   

As previously shown in Figure~\ref{fig_jetcorr}, there is strong evidence 
from STAR and PHENIX that partons propagating through the medium have a 
fragmentation pattern showing an increased yield of hadrons at low $p_T$ 
and a very broad angular distribution.  The possible two-peaked structure 
in the angular distribution and the correlation of this structure with 
modification in the $<p_{T}>$ have led to much speculation about the 
response of the medium to the deposited energy.  It is possible 
that this response takes the form of a Mach cone of sound waves
\cite{Casalderrey-Solana:2004qm,Satarov:2005mv}, or a wake in the 
color fields \cite{Ruppert:2005uz}, or is the result of gluonic 
Cherenkov radiation due to an elevated index of (gluon) refraction 
of the medium \cite{Dremin:2005an,Koch:2005sx}.  The detailed
measurement of identified multi-particle jet correlations may help 
distinguish between these scenarios. These analyses require very high 
statistics which will be available after the luminosity upgrade.

The heavy quarkonia program at RHIC has just started.  Additional upgrades 
to detectors and improvements in luminosity are needed to fully quantify 
the level of suppression in heavy ion reactions relative to proton-proton 
and proton-gold or deuteron-gold reactions over a wide kinematic range.  
In addition, measurements of the different quarkonia states are important 
for determining whether 
the suppression relates to color screening or to collisional interactions 
with other partons.  The lattice results for spectral functions in the 
quenched approximation suggest that  measurements of the less strongly 
bound $\chi_{c}$ states should be a focus of the program.  The deeply 
bound $\Upsilon(1s)$ is also interesting, because it may be unaffected by 
the medium if the temperature reached at RHIC is not much above $T_{c}$.  
Additional data on open charm yields and flow will help confirm 
or refute models predicting large $J/\psi$ formation from initially 
uncorrelated $c$ and $\overline{c}$ via recombination.

Recent preliminary results from the NA60 experiment~\cite{na60} at the CERN-SPS have 
confirmed the earlier NA45 results~\cite{na45} of a low-mass enhancement in dilepton 
production below the $\rho(770)$ mass.  The excellent resolution and 
precision of these new results has reignited hopes that measurements 
of the dilepton spectrum will permit a determination of the modification
of the $\rho$ spectral function in the medium and, indirectly, yield 
evidence for the gradual restoration of chiral symmetry upon approach
to $T_c$.  These measurements are extremely challenging at RHIC as the 
backgrounds are very large.  Upgrades for the PHENIX experiment with a 
hadron blind detector and improved Dalitz rejection using time of flight 
and micro-vertex detection in the STAR experiment will make these 
measurements possible.

One can also measure thermal radiation with direct or virtual photons.  
Again, there is large background from $\pi^{0}$ and $\eta$ decays, but 
the rewards for successful efforts are great.  
A measurement of the thermal temperature of the medium is a 
critical step towards detailing the number of active degrees of freedom in the 
equation of state.  PHENIX has recently introduced a new analysis technique 
for heavy ions by measuring virtual photons via $e^{+}e^{-}$ pairs and 
then subtracting the hadron decay contributions by studying the invariant 
mass distribution of the pairs~\cite{henner}.  Additionally the measurement 
of $\gamma\gamma$ HBT correlations~\cite{gghbt} to eliminate these decay 
contributions may help to constrain the time evolution of the temperature.

%% file: section4b.tex
In the very near future, the Large Hadron Collider (LHC) at CERN will be commissioned for 
high energy proton-proton and heavy ion lead-lead reactions.  These heavy 
ion collisions will be at a center-of-mass energy of $5.5~TeV$ per nucleon 
pair, thus bringing $1144~TeV$ in total kinetic energy into collision.  
A detailed overview of the program is given in ~\cite{lhc_yellow}.
There are three experiments that will collect data during heavy ion running:  
ALICE, ATLAS and CMS.  The ALICE experiment is a detector system dedicated 
to heavy ion physics with capabilities designed for the characterization 
of the quark-gluon plasma.  The requirements include large coverage to low 
transverse momentum, excellent particle identification, and specialized 
detectors for specific channels of interest (for example $J/\psi, 
\Upsilon \rightarrow l^{+}l^{-}$).  ATLAS and CMS are detectors primarily 
designed for discovering the Higgs boson, supersymmetric particles, and 
other new physics.  Both detectors have excellent calorimetry with large 
coverage and inner silicon detectors for charged particle tracking.  
Thus, the experimental program at the LHC is well equipped, and despite 
only one month of heavy ion running per year, the large rate capabilities 
of these detectors make the program very promising.

It is possible that at the LHC one will achieve energy densities such 
that a weakly interacting quark-gluon plasma is formed, because thermal 
interactions are of sufficiently large $q^{2}$ that $\alpha_s \ll 1$.  
This would be extremely interesting as one would see the perturbative
calculations which have failed to describe data at RHIC then becoming 
applicable at the LHC.  Such a scenario could imply, for instance, that 
the elliptic flow $v_2$ at LHC may be smaller than at RHIC energies.
It is also possible that the quark-gluon plasma at LHC energies will 
still be a strongly interacting system for most of the medium's time 
evolution.

Another exciting part of the program at the LHC is that the gluon density 
in the incoming nuclei is much higher than at RHIC.  Thus, studies of 
saturation physics will enter a new regime at the LHC.  Additionally, 
measurements of $\gamma+jet$ events at forward angles will allow precise 
kinematic determination of the initiating partons ($x_{1},x_{2}$), and 
thus will allow for precise measurements of deviations from predictions 
based on factorized parton distributions.  

The higher energies of the LHC not only imply a higher initial 
temperature, but also that even higher $q^{2}$ 
probes of the medium are available and abundantly produced.  Quark and 
gluon jets will be measureable well above $p_{T} > 200~GeV$.  In addition, 
the tagging of heavy flavor jets from displaced vertices is possible 
in all three experiments.  At these very high $p_T$ the interactions 
with the medium are expected to be completely dominated by induced gluon 
radiation and describable by pQCD.   Quarkonia studies will also benefit 
from the higher energy as abundant production of the $\Upsilon (1s,2s,3s)$ 
states will allow detailed studies of these bound states.

The RHIC and LHC programs should prove to be quite complementary whether 
it turns out that LHC induced collisions are in the strongly coupled or 
weakly coupled regimes or somewhere in between.

%% file: section4d.tex

The main unsolved theoretical problems concerning nuclear collisions at RHIC 
fall into five classes:
\begin{enumerate}
\item
Understanding the mechanisms behind early equilibration and thermalization,
and entropy production, in general;
\item
The microscopic structure of the quark-gluon plasma in the temperature 
region above, but not far above, $T_c$;
\item
The numerical solution of three-dimensional, viscous relativistic 
fluid dynamics;
\item
The effect of localized energy deposition by a hard parton on the medium.
\item
The process of bulk hadronization.
\end{enumerate}
This does not mean that there are not many other interesting and important
theoretical problems, but these generally can be approached in a now well
established framework. What distinguishes the five issues listed above is
that the framework, in which they can be addressed, is still uncertain.
Thus, all we can do here is to describe some ideas that hold promise
for future studies.

{\em Equilibration:} The most promising conceptual framework for a rigorous
treatment of parton equilibration on the basis of QCD currently is the 
idea that nuclear matter is transformed in the collision from saturated 
gluonic matter in a universal quantum state (the color glass condensate) 
to a thermalized and equilibrated quark-gluon plasma \cite{McLerran:2005kk}. 
The theoretical arguments for this scenario invoke asymptotically high beam 
energies; whether this condition already applies in the energy domain at 
RHIC, remains to be seen. Three steps can be distinguished in this process: 
the decoherence of the initial quasi-classical quantum state into 
particle-like quanta; the approach to local isotropy of the resulting 
parton distribution; finally, thermalization and chemical equilibration.

The decoherence of the initial state has so far been studied mostly by
means of numerical solution of the Yang-Mills equations
\cite{Krasnitz:2002mn,Krasnitz:2003jw,Lappi:2003bi}. These simulations 
do not, however, answer the question how entropy is created in this process. 
A recent attempt to estimate the decoherence time by analytical techniques
gives a value of the order of $Q_s^{-1} \sim 0.2$~fm/c \cite{Muller:2005yu}. 
The entropy created solely by the decoherence (``shattering'') of the 
color glass condensate can be estimated to be about 1/3 of the finally 
observed entropy \cite{Muller:2003cr}.

A promising mechanism to explain the apparent early isotropization of 
the distribution of liberated partons are the instabilities which 
generically exist in a parton plasma with an anisotropic momentum 
distribution \cite{Mrowczynski:2005ki}. These lead to the exponential 
growth of coherent color fields, dominantly of color-magnetic nature, 
which can efficiently deflect partons. It is not clear at present 
whether these fields can also speed up complete thermalization 
\cite{Mueller:2005un}, for example by enhancing gluon radiation via 
synchrotron-like emission of soft quanta \cite{Shuryak:2002ai}. Since the 
longitudinal expansion of the matter, which persists even at late times,
continually acts as a driver of a momentum space anisotropy in the 
parton distribution, coherent fields due to plasma instabilities may 
well remain present in the medium at all times, as long as the temperature
exceeds $T_c$. They may thus be a part of the structure of the matter
that we are studying experimentally, rather than just a mechanism for
its equilibration.

{\em Structure of the quark-gluon plasma:} A detailed insight into 
the microscopic structure of thermal QCD matter in the region 
$T_c \leq T \leq 2T_c$ is of utmost importance for our understanding
of the physics of most experimental probes. In principle, the tool 
for the theoretical investigation of this structure exists: lattice 
gauge theory. A problem is that it is not sufficient to calculate 
various quantities on the lattice; one must also be able to extract
a physical picture from the lattice results. Such studies usually 
proceed in various levels of sophistication: from global properties, such
as the equation of state, over various susceptibilities describing 
fluctuations around the equilibrium state and static correlation functions,
to dynamical correlations and spectral functions. High quality lattice 
results now exist for the equation of state, susceptibilities and many 
static correlators, but a systematic analysis of these results in terms
of physical models is still in progress. The calculations of spectral 
functions on the lattice are still in their infancy and presently 
restricted to calculations without dynamical quarks. This greatly 
limits their usefulness at the present time.

The recent progress in solving strongly coupled supersymmetric gauge
theories by means of duality techniques is another very promising approach
(see \cite{Klebanov:2005mh} for a recent review).
One problem here is that the connection between the degrees of freedom 
in the gauge theory at weak and strong coupling is not well understood,
making it difficult to model the transition between the two regimes.
Another hurdle is that it is difficult to formulate supersymmetric gauge 
theories on the lattice \cite{Catterall:2005df} and study them across the 
entire range of coupling constants. Another possible approach to study the 
structure of the quark-gluon plasma away from weak coupling is molecular 
dynamics \cite{Hartmann:2006nb,Gelman:2006xw}.

{\em Viscous 3-D fluid dynamics:} The RHIC data clearly demonstrate that
the medium produced in nuclear collisions is not boost invariant. A 
realistic treatment of the reaction therefore requires full solutions 
of relativistic fluid dynamics in three dimensions. Such solutions 
have recently become available, including hadronic cascade treatments 
of the break-up phase \cite{Hirano:2002ds,Hama:2005dz,Nonaka:2005aj}.
An important missing aspect of these solutions is the presence of a
physical viscosity. Although the data suggest that the viscosity should
be small in the plasma phase, full three-dimensional (numerical) solutions 
of viscous, relativistic hydrodynamics for a given equation of state and
realistic initial conditions are needed in order to deduce an upper limit 
of the viscosity from the collective flow data.

{\em Medium response to energy deposition:} The current interest in this
problem is driven by the experimental observations of an enhancement of
soft hadron emission over a wide angular range in association with a
recoiling energetic hadron. The theoretical problem here is presumably 
quite well defined: What is the response of the thermal medium to a 
color charge that propagates through it at nearly the speed of light
\cite{Ruppert:2005uz} ? The problem is complicated by the fact that 
the color charge itself evolves on its way through the plasma due to 
scattering and radiation. It is an interesting question whether a scale 
exists that separates the evolution of the propagating charge from the 
response of the medium. Of course, one needs to bear in mind that, even 
if such a scale exists in principle, the energies accessible at RHIC may 
not be sufficiently high for its application. Present models suggest that 
an explanation of the data in terms of a Mach 
\cite{Casalderrey-Solana:2004qm,Ruppert:2005uz,Satarov:2005mv}
or Cherenkov \cite{Dremin:2005an,Koch:2005sx} cone requires an extremely
efficient coupling between the color field of the penetrating parton 
amd the collective modes of the medium \cite{Renk:2005si}. Studies of 
possible mechanisms for this coupling, either by perturbative techniques 
or molecular dynamics simulations, will be important.

{\em Bulk hadronization:} The competition of hadronization in bulk 
with the fragmentation mechanism familiar from jet formation was 
proposed early on as a signature of the formation of a deconfined 
quark-gluon plasma \cite{Lopez:1984st}. But before the advent of RHIC, 
the kinematic range of transverse momenta $p_T$ where this competition 
can be studied was inaccessible. As discussed in Section~\ref{sect3d}, the 
thermal recombination model explains some of the more surprising features of 
hadron production observed in the RHIC experiments. The persistence of 
dihadron correlations \cite{Adler:2002tq,Adler:2004zd} in the $p_T$-domain 
where recombination appears to dominate over vacuum fragmentation 
indicates that either recombination processes involving both, hard 
scattered and thermal partons \cite{Hwa:2004ng}, or local momentum 
correlations in the hadronizing medium \cite{Fries:2004hd} are important. 
This suggests the need for a theoretical framework that treats fragmentation 
and recombination as two limiting cases of a unified hadronization scheme. 
First steps in this direction have recently been taken \cite{Majumder:2005jy}.

%% file: section5.tex

Relativistic heavy ion physics has witnessed several paradigm shifts
during the past decade driven largely by experimental discoveries. 
The power of large, multipurpose detectors suited to the simultaneous 
measurement of many observables has also been demonstrated. As a result 
of these innovations and the availability of a dedicated accelerator 
facility, RHIC, much progress has been made in a relatively short time.

In Section \ref{sect1b} we stated that the experimental quest for the 
quark-gluon plasma must proceed in steps: First, the determination
that the particles produced in the nuclear reaction really form, for a 
brief period, matter that deserves a description in thermodynamic terms; 
second, the determination that this matter has a novel structure and is 
not just a dense gas of hadrons; and third, the characterization of its 
main physical properties. At the time of this review, the first two steps
have clearly been achieved. We are confident that the matter has been 
thermodynamically equilibrated, probably at an early time in the reaction,
and we have found compelling evidence for the case that the structure of
this matter is different from any known before. The third step, involving
the characterization of its properties by systematic experimental 
investigation, is in progress. In order to succeed, it will require 
improvements in the RHIC detectors and a substantial increase in the
available luminosity.

On the theory side, the original expectation that the quark-gluon plasma
would be a ``simple'' state of matter characterized by largely 
perturbative interactions and transport processes has fallen by the 
wayside. It has been replaced with the understanding that the quark-gluon 
plasma, at least in the parameter regime accessible at RHIC, is a state 
of matter characterized by strong interactions among its constituents and
by novel properties, such as a very low viscosity and large stopping power.
This insight is not entirely new; the temperature region near $T_c$ was
always recognized as one for which calculations are hard and quantitative 
predictions dificult. The results from RHIC have forced theorists to 
face these difficulties.

No consensus has yet emerged about what the microscopic structure of
this new state is, but we do have an ever improving tool - lattice 
gauge theory - that allows us to determine some of its properties in 
a model independent way. One of the results obtained in this way is 
the equation of state and the recognition that it does not contain a
strong discontinuity, such as a first-order phase transition, under
the conditions (approximately vanishing net baryon number density)
prevailing at RHIC. This insight, in turn, has done away with the 
previously held expectations of discontinuities in experimental data
as function of beam energy, impact parameter, or nuclear size. Any 
changes observed as function of these parameters must be continuous,
even though it may be rapid, as the transition from hadronic matter 
to a quark-gluon plasma takes place. 

Perhaps the most basic insight from the RHIC data is that nearly 
thermalized matter is, indeed, produced in nuclear collisions. In fact,
the ensemble of emitted hadrons is so well thermalized, both kinetically
and chemically, that one has to look at rare probes, such as unstable 
resonances or hadrons emitted at high transverse momentum, to find 
deviations from equilibrium. 
These results leave us quite confident that matter, in the thermodynamic
sense, is being produced. The large collective transverse flow, which
shows a quadrupole anisotropy in noncentral collisions, signals that 
the equilibration happens very early in the reaction, probably at time
less than 1 fm/c. This conclusion relies on comparisons with solutions 
of relativistic fluid dynamics, which can be made to agree well with the 
data for realistic equations of state. Calculations using viscous fluid
dynamics, with certain somewhat questionable approximations, lead to 
an even stronger conclusion: that the matter must be nearly inviscid, 
a nearly ``perfect'' fluid, and the mean free paths of particles in it
must be very short.

The other important new insight is that the medium produced in nuclear
collisions has a large opacity to energetic particles carrying a color
charge. The numbers extracted from the RHIC data push the limits of 
possible explanations within the perturbative theory of energy loss 
for light partons in QCD, and they clearly exceed them for heavy quarks.
Again, the mean free path must be shorter than expected in a perturbative 
framework. The stopping power of the medium makes much more detailed
measurements of jet quenching possible at RHIC than was anticipated, 
and ``jet tomography'' has become an important experimental tool for 
the characterization of the medium, its geometry, and its evolution.

This summary would be incomplete without mentioning that not all data 
measured in nuclear collisions are well explained and understood. A 
prime example are the measurements of identical particle correlation
functions, which do not agree with the predictions of fluid dynamical
calculations that fit the hadron spectra. It may well be that the 
solution of this ``HBT puzzle'' simply requires more realistic 
solutions of fluid dynamics which include a full three-dimensional 
evolution and viscosity.  Some other tantalizing features of the data, 
such as strong angular correlations among hadrons in the same momentum 
range, have not yet received a generally accepted explanation but have
generated various intriguing speculations.

The many remaining questions about the nature of the matter that has
been discovered at RHIC pose great challenges to experiment and 
theory. On the theoretical side, we need to develop tools that permit
a systematic calculation of the structure of thermal QCD matter and
its transport properties when the coupling is not weak. The string 
theory inspired techniques now used to study supersymmetric gauge 
theories at strong coupling promise to be useful also for real QCD. 
More conventional tools developed for the description of turbulent 
electromagnetic plasmas will perhaps allow us to solve the problem 
of equilibration.

On the experimental side, more measurements of rare probes which 
escape complete thermalization, such as heavy quarks, photons, and
lepton pairs will be essential. Another important avenue for future 
studies are multi-particle correlations, which can give information 
about the medium response to a hard QCD process. Both kinds of 
measurements require significant detector improvements and an 
increase in the collider luminosity. Our capability to execute such
a program will grow significantly with the RHIC luminosity upgrade
and with the start of the heavy ion experiments at the Large Hadron 
Collider. A detailed set of measurements needed to guide theory and to 
discriminate between different theoretical approaches is now being 
planned.